\documentstyle[12pt]{article}

\begin{document}

\setcounter{page}{1}

\pagestyle{plain} \vspace{1cm}
\begin{center}
\Large{\bf Gauss-Bonnet Cosmology with Induced Gravity and Non-Minimally Coupled
Scalar Field on the Brane}\\
\small \vspace{1cm} {\bf Kourosh Nozari $^{a,b,}$ \footnote{
knozari@umz.ac.ir}}\quad and \quad {\bf Behnaz Fazlpour $^{a,}$
\footnote{
b.fazlpour@umz.ac.ir}}\\
\vspace{0.5cm} {\it $^{a}$Department of Physics, Faculty of Basic
Sciences,\\
University of Mazandaran,\\
P. O. Box 47416-1467, Babolsar, IRAN}\\
{\it $^{b}$ Research Institute for Astronomy and Astrophysics of
Maragha, \\P. O. Box 55134-441, Maragha, IRAN}

\end{center}
\vspace{1.5cm}
\begin{abstract}
We construct a cosmological model with non-minimally coupled scalar
field on the brane, where Gauss-Bonnet and Induced Gravity effects
are taken into account. This model has $5D$ character at both high
and low energy limits but reduces to $4D$ gravity in intermediate
scales. While induced gravity is a manifestation of the IR limit of
the model, Gauss-Bonnet term and non-minimal coupling of scalar
field and induced gravity are essentially related to UV limit of the
scenario. We study cosmological implications of this scenario
focusing on the late-time behavior of the solutions. In this setup,
non-minimal coupling plays the role of an additional fine-tuning
parameter that controls the initial density of predicted finite
density big bang. Also, non-minimal coupling has important
implication on the bouncing nature of the solutions.\\
{\bf PACS}: 04.50.+h,\, 98.80.-k, 98.80.Cq\\
{\bf Key Words}: Scalar-Tensor Gravity, Induced Gravity, Stringy
Effects, Cosmological dynamics
\end{abstract}
\newpage
\section{Introduction}
The idea that we and all standard matter live on a brane embedded in
a higher dimensional bulk has attracted a lot of attention[1-3]. In
this view point, extra dimensions are accessible only for graviton
and possibly non-standard matter. The setup of Randall and Sundrum
(RSII) considers the observable universe as a $3$-brane with
positive tension embedded in five dimensional anti-de sitter bulk.
In the low energy limit, the $5D$ graviton is localized on the brane
due to the warped geometry of the bulk. The notion of AdS/CFT
correspondence help us to understand this property since the
$4D$ gravity is coupled to a conformal field in the RS model [4,5].\\
The effect of the bulk on the brane can be determined by the
effective mass $\mathcal{M}$ of the bulk fluid that is measured by a
bulk observer at the brane. For spherically symmetric brane, this
mass can be considered as effective gravitational mass of the bulk.
This mass depends on the brane scale factor and the proper time on
the brane. In the case that the bulk observer is comoving with the
bulk fluid, the mass is assumed to be comoving. However, for matter
components such as a bulk radiation fluid, there is no comoving
observer[6,7].

On the other hand, the model proposed by Dvali, Gabadadze and
Porrati (DGP) is a radiative correction, the bulk is a flat
Minkowski spacetime, but a reduced gravity term appears on the brane
without tension. This model is different in this respect that it
predicts deviations from the standard $4$-dimensional gravity over
large distances. In this scenario, the transition between four and
higher-dimensional gravitational potentials arises due to the
presence of both the brane and bulk Einstein terms in the action.
Existence of a higher dimensional embedding space allows for the
existence of bulk or brane matter which can certainly influence the
cosmological evolution on the brane. This model has a rich
phenomenology discussed in [8].  Maeda, Mizuno and Torii have
constructed a braneworld scenario which combines the Randall-Sundrum
II ( RSII) model and DGP model[9]. In this combination, an induced
curvature term appears on the brane in the RSII model. This model
has been called {\it warped DGP braneworld} in literature[10]. The
existence of induced gravity term leads to a self-accelerating
branch in the brane evolution[11,12].\\ Braneworld model with scalar
field minimally or non-minimally coupled to gravity have been
studied extensively(see[13] and references therein). The
introduction of non-minimal coupling (NMC) is not just a matter of
taste; it is forced upon us in many situations of physical and
cosmological interests. For instance, NMC arises at the quantum
level when quantum corrections to the scalar field theory are
considered. Even if for the classical, unperturbed theory this NMC
vanishes, it is necessary for the renormalizability of the scalar
field theory in curved space. In most theories used to describe
inflationary scenarios, it turns out that a non-vanishing value of
the coupling constant cannot be avoided. In general relativity, and
in all other metric theories of gravity in which the scalar field is
not part of the gravitational sector, the coupling constant
necessarily assumes the value of \, $\frac{1}{6}$\,. The study of
the asymptotically free theories in an external gravitational field
with a Gauss-Bonnet term shows a scale dependent coupling parameter.
Asymptotically free grand unified theories have a non-minimal
coupling depending on a renormalization group parameter that
converges to the value of $\frac{1}{6}$ or to any other initial
conditions depending on the gauge group and on the matter content of
the theory. An exact renormalization group study of the
$\lambda\phi^4$ theory shows that NMC$=\frac{1}{6}$ is a stable
infrared fixed point. Also in the large $N$ limit of the
Nambu-Jona-Lasinio model, we have NMC$=\frac{1}{6}$. In the $O(N)$-
symmetric model with $V =\lambda\phi^4$, NMC  is generally nonzero
and depends on the coupling constants of the individual bosonic
components. Higgs fields in the standard model have NMC$=0$ or
$\frac{1}{6}$. Only a few investigations produce zero value(for a
more complete discussion of these issues we refer to papers by V.
Faraoni, specially Ref. [14] and references therein). In view of the
above results, it is then natural to incorporate an explicit NMC
between scalar field and induced Ricci scalar on the brane.

On the other hand, in a braneworld scenario, the radiative
corrections in the bulk lead to higher curvature terms. At high
energies, the Einstein-Hilbert action will acquire quantum
corrections. The Gauss-Bonnet (GB) combination arises as the leading
bulk correction in the case of the heterotic string theory [15].
This term leads to second-order gravitational field equations linear
in the second derivatives in the bulk metric which is ghost free
[16-18], the property of curvature invariant of the Gauss-Bonnet
term.\\
Inclusion of Gauss-Bonnet term in the action results in a variety of
novel phenomena which certainly affects the cosmological
consequences of these generalized braneworld setup, although these
corrections are smaller than the usual Einstein-Hilbert terms
[19-21]. Moreover, the zero mode of graviton has been localized in
the GB model [22]. The cosmological evolution corresponding to RS
model in the presence of a bulk GB term has been considered in
[17,23-28] see also [29]. Also the case of minimally coupled scalar
field with
GB gravity has been discussed extensively[30-33].\\
In the presence of GB term with induced gravity, there are different
cosmological scenarios, even if there isn't any matter in the
bulk[34]. In this paper, we generalize the previous studies to the
case that a scalar field non-minimally coupled to induced curvature
is present on the brane in the presence of radiative corrections. We
first review briefly the setup of
Brown-Maartens-Papantonopoulos-Zamarias(BMPZ)[35]. Then we
generalize this setup to the more general framework of scalar-tensor
theories. We show that relative to BMPZ scenario, there are several
interesting features which affect certainly the cosmological
dynamics on the brane. Since Gauss-Bonnet and induced gravity
effects are related to two extremes of the scenario (UV and IR
limits), inclusion of sringy effects via Gauss-Bonnet term leads to
a finite density big bang[35]. This interesting feature has been
explained in a fascinating manner by $T$-duality of string
theory[36]. On the other hand, non-minimal coupling itself accounts
for a non-singular soft big bang scenario[37]. In our setup,
existence of non-minimal coupling of scalar field and induced
gravity on the brane, controls the initial density of this finite
big bang scenario. In other words, in this framework, non-minimal
coupling with its special fine-tuning ( see [38] and references
therein), plays the role of a parameter that can control density of
matter fields at the beginning of the universe. As we will show,
incorporation of both GB and non-minimal coupling effects will
enhance special characters of BMPZ scenario.

\section{DGP Inspired Scalar-Tensor Theories}
The action of the DGP scenario in the presence of a non-minimally
coupled scalar field on the brane can be written as follows [39]
$$S=\frac{1}{2\kappa_{5}^{2}}\int d^{5}x\sqrt{-g^{(5)}}\Big[
R^{(5)}-2\Lambda_{5}\Big]$$
\begin{equation}
+\Bigg[\frac{r}{2\kappa_{5}^{2}}\int
d^{4}x\sqrt{-g}\bigg(\alpha(\phi) R -2\kappa_{4}^{2} g^{\mu\nu}
\nabla_{\mu}\phi\nabla_{\nu}\phi -4\kappa_{4}^{2}V(\phi) -4
\kappa_{4}^{2}\lambda\bigg)\Bigg]_{y=0},
\end{equation}
where we have included a general non-minimal coupling $\alpha(\phi)$
in the brane part of the action. $y$ is coordinate of the fifth
dimension and we assume the brane is located at $y=0$.\,
$g^{(5)}_{AB}$ is five dimensional bulk metric with Ricci scalar
${R^{(5)}}$, while $g_{\mu\nu}$ is induced metric on the brane with
induced Ricci scalar $R$.\, $g_{AB}$ and $g_{\mu\nu}$ are related
via $g_{\mu\nu}={\delta_{\mu}}^{A}{\delta_{\nu}}^{B}g_{AB}$.
$\lambda$ is the brane tension (constant energy density) and $r$ is
the cross-over scale that is defined as follows
\begin{equation}
r=\frac{\kappa_{5}^{2}}{2 \kappa_{4}^{2}}=\frac{M_{4}^{2}}{2
M_{5}^{3}}.
\end{equation}
The generalized cosmological dynamics of this setup is given by the
following Friedmann equation [34,40]
\begin{equation}
\varepsilon\sqrt{H^{2}-\frac{
\Upsilon}{a^{4}}-\frac{\Lambda_{5}}{6}+\frac{K}{a^{2}}}=r
\alpha(\phi)\Big(H^{2}+\frac{K}{a^{2}}\Big)-
\frac{\kappa_{5}^{2}}{6}(\rho+ \rho_{\phi}+\lambda).
\end{equation}
where $\varepsilon=\pm 1$ is corresponding to two possible branches
of DGP cosmology and $\Upsilon$ is the bulk black hole mass which is
related to the bulk Weyl tensor. This mass, as generalized dark
radiation, induces mirage effects in the evolution and the
gravitational effect of the bulk matter on the brane evolution can
be described in terms of this mass as measured by a bulk observer at
the location of the brane (the DGP limit has a Minkowski bulk
$\Lambda_{5}=0$ with $\Upsilon=0$). A part of the effects of
non-minimal coupling of scalar field $\phi$ with gravity is hidden
in the definition of the effective energy density. Assuming the
following line element
$$ds^{2}=q_{\mu\nu}dx^{\mu}dx^{\nu}+b^{2}(y,t)dy^{2}=-n^{2}(y,t)dt^{2}+
a^{2}(y,t)\gamma_{ij}dx^{i}dx^{j}+b^{2}(y,t)dy^{2},$$ where
$\gamma_{ij}$ is a maximally symmetric 3-dimensional metric defined
as $\gamma_{ij}=\delta_{ij}+k\frac{x_{i}x_{j}}{1-kr^{2}}$,  the
energy density of non-minimally coupled scalar field on the brane is
given as follows [39,41]
\begin{equation}
\rho_{\phi}=\left[\frac{1}{2}\dot{\phi}^{2}+n^{2}V(\phi)-6\alpha'H\dot{\phi}\right]_{y=0},
\end{equation}
where \,$H=\frac{\dot{a}}{a}$\, is Hubble parameter,\,
$\alpha'=\frac{d\alpha}{d\phi}$ and $\dot{\phi}=\frac{d\phi}{dt}$.

If we consider a flat brane (K=0) with $\lambda=0$ and also a
Minkowski bulk ($\Lambda_{5}=0$, $\Upsilon=0$), then we can write
equation (3) as follows
\begin{equation}
H^{2}=\pm \frac{H}{r
\alpha(\phi)}+\frac{\kappa_{4}^{2}}{3\alpha(\phi)}(\rho
+\rho_{0\phi}-6\alpha'H\dot{\phi}).
\end{equation}
where
$\rho_{0\phi}=\left[\frac{1}{2}\dot{\phi}^{2}+n^{2}V(\phi)\right]_{y=0}$.
The DGP model has two branches, i.e \, $\varepsilon=\pm 1$
corresponding to two different embedding of the brane in the bulk.
The behavior of two branches at high energies and low energies are
summarized as follows: \\
In the high energy limit we find
\begin{equation}
\hspace{1.5
cm}DGP(\pm):\,\,\,\,\,\,\,\,H^{2}=\frac{\kappa_{4}^{2}}{3\alpha(\phi)}(\rho
+\rho_{\phi}),
\end{equation}
while in low energy limit we have \\
$$ DGP(+):\,\,\,\,\,\,\,\,H\longrightarrow \frac{1}{r
\alpha(\phi)}-2\frac{\kappa_{4}^{2}\alpha'\dot{\phi}}{\alpha(\phi)}$$
\begin{equation}
DGP(-):\,\,\,\,\,\,\,\,H=0.
\end{equation}
In terms of dimensionless variables introduced in [35]
\begin{equation}
h=Hr,\,\,\,\mu=\frac{r\kappa_{5}^{2}}{6}\rho,\,\,\,\,
\mu'=\frac{r\kappa_{5}^{2}}{6}\rho_{0\phi},\,\,\,\,\sigma=\frac{r\kappa_{5}^{2}}{6}\lambda
,\,\,\,\,\tau=\frac{t}{r},
\end{equation}
we find
\begin{equation}
h^{2}=\pm \frac{h}{
\alpha(\phi)}+\frac{(\mu+\mu')}{\alpha(\phi)}-\frac{2 h
\kappa_{4}^{2}}{\alpha(\phi)}\frac{d \alpha}{d \tau}.
\end{equation}
The solutions of this equation for $h$ are as follows
\begin{equation}
h=\pm\frac{1}{2\alpha(\phi)}-\frac{\kappa_{4}^{2}}{\alpha(\phi)}\frac{d
\alpha}{d \tau}+\frac{\sqrt{1\mp 4\kappa_{4}^{2}\frac{d \alpha}{d
\tau}+4(\kappa_{4}^{2}\frac{d \alpha}{d
\tau})^{2}+4\alpha(\phi)(\mu+\mu')}}{2\alpha(\phi)}.
\end{equation}
Here the negative root is not suitable since in the limit of \, $\mu
+\mu'\longrightarrow 0$, with this sign one cannot recover the low
energy limit of the model highlighted in (7).
\begin{figure}[htp]
\begin{center}
\includegraphics{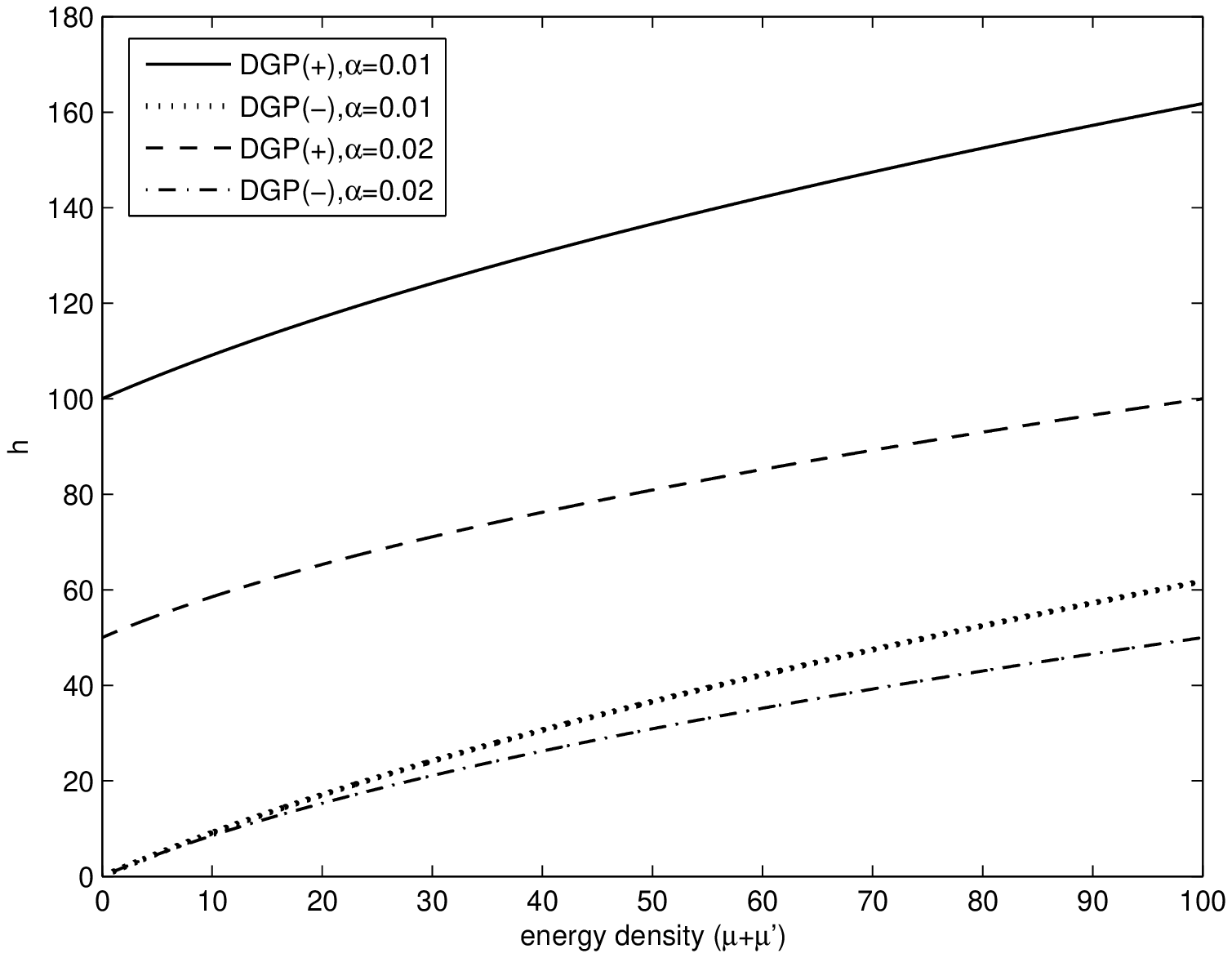}
\end{center}
\vspace{7 cm}
 \caption{\small { Two possible branches of DGP-inspired non-minimal model. The non-minimal
 coupling of scalar field is assumed to be positive and the brane is considered to be tensionless, $\sigma=0$.}}
\end{figure}

\begin{figure}[htp]
\begin{center}
\includegraphics{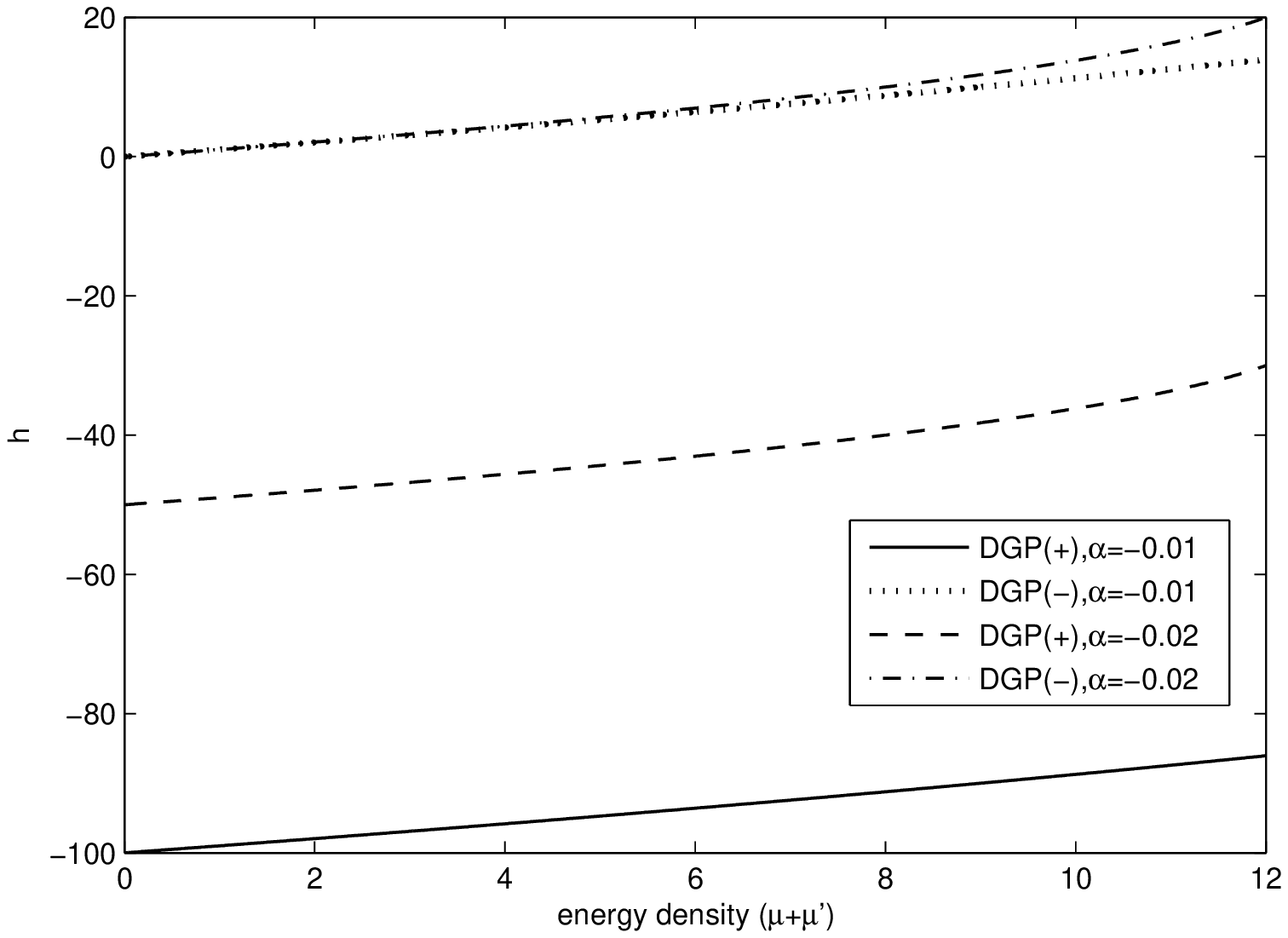}
\end{center}
\vspace{5 cm}
 \caption{\small { Two possible branches of DGP-inspired non-minimal model. The non-minimal
 coupling of scalar field is assumed to be negative and the brane is considered to be tensionless, $\sigma=0$.}}
\end{figure}
Figure $1$ shows the behavior of these solutions with some specific
values of non-minimal coupling\footnote{To plot all of the figures
in this paper, equation (41) acts as a condition on the values that
$\alpha$ can attains. The possible values of $\gamma$ extracted from
observational data are shown in table (1). Using SNIa+LSS+H(z) test,
we obtain $\alpha\geq 0.01$ and $\alpha\leq -0.01$ approximately.
For simplicity in drawing figures we have assumed that scalar field
has no dynamics, i.e. $\frac{d\phi}{d\tau}=0$. The time dependent
non-minimal coupling will be discussed at the end of the paper. }.
The upper sign in relation (10) is related to DGP(+) and lower sign
for DGP(-). It is seen that there is a late-time self-acceleration
in the DGP(+) branch similar to the minimal case, however, in
minimal case when $\mu\longrightarrow0$ then $h\longrightarrow1$,
whereas in the non-minimal case, in this limit i.e \,
$\mu+\mu'\longrightarrow0$, we have
$h\longrightarrow\frac{1}{\alpha(\phi)}-2\frac{\kappa_{4}^{2}}{\alpha(\phi)}\frac{d
\alpha}{d \tau}$. In DGP(+) branch, the endpoint is a vacuum de
Sitter state and Anti de Sitter state for positive and negative
non-minimal coupling respectively whereas in the DGP(-) branch, the
endpoint is a Minkowski state. The effect of non-minimal coupling in
this case is to shift the end point of DGP(+) branch. Depending on
the value that $\alpha(\phi)$ can attain [38], the late-time
acceleration of the universe can be fine-tuned properly.

Now for $\lambda\neq0$, the solutions of dimensionless Friedmann
equation are as follows
\begin{equation}
h=\pm\frac{1}{2\alpha(\phi)}-\frac{\kappa_{4}^{2}}{\alpha(\phi)}\frac{d
\alpha}{d \tau}+\frac{\sqrt{1\mp 4\kappa_{4}^{2}\frac{d \alpha}{d
\tau}+4(\kappa_{4}^{2}\frac{d \alpha}{d
\tau})^{2}+4\alpha(\phi)(\mu+\mu'+\sigma)}}{2\alpha(\phi)}.
\end{equation}
These solutions are shown in figures $3$ and $4$ with $\sigma\neq
0$. For negative tension in DGP(+) branch, the endpoint is a vacuum
de Sitter state and there is a self-acceleration whereas in DGP(-)
branch the solutions terminate at finite density (fig.3). For
positive tension, both of the solutions $(DGP(\pm))$ have
self-acceleration and the
endpoints are vacuum de Sitter state (fig.4).\\
The existence of the energy density $(\lambda)$ on the brane gives
rise to a shift of the solutions. Moreover, existence of
$\alpha(\phi)$ leads to further shift of these solutions.\\
\begin{figure}[htp]
\begin{center}
\includegraphics{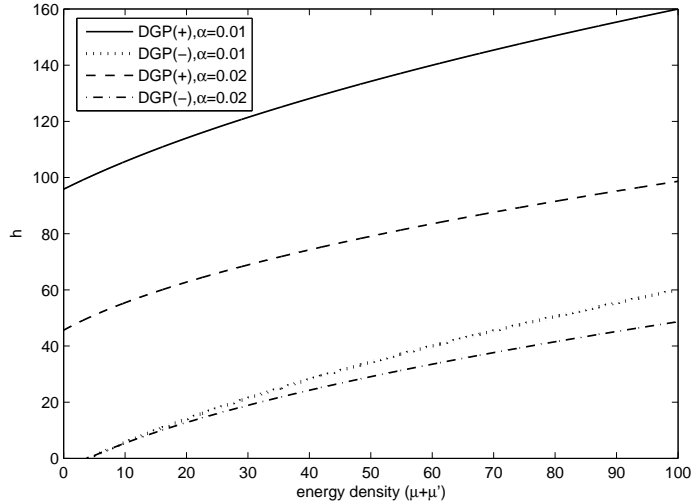}
\end{center}
\vspace{6 cm}
 \caption{\small { Two possible branches of DGP inspired non-minimal model with negative tension
 brane and positive non-minimal coupling. We have set $\alpha(\phi)=0.01, 0.02$ and $\sigma=-4$.}}
\end{figure}
\begin{figure}[htp]
\begin{center}
\includegraphics{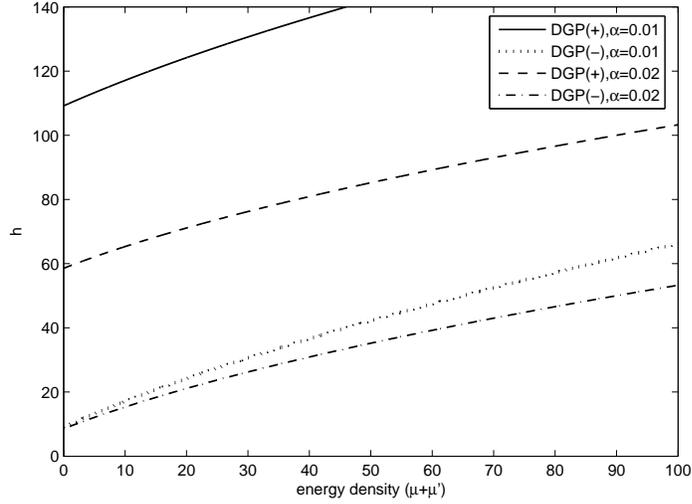}
\end{center}
\vspace{5 cm}
 \caption{\small { Two possible branches of DGP inspired non-minimal model with positive
 brane tension and positive non-minimal coupling. We have set $\alpha(\phi)=0.01, 0.02$ and $\sigma=10$.}}
\end{figure}\\
DGP model is an IR modification of general relativity. In the UV
limit, stringy effects will play important role. In this viewpoint,
to discuss both UV and IR limit of the scenario simultaneously, the
DGP model is not sufficient and we should incorporate stringy
effects via inclusion of the Gauss-Bonnet terms.
\section{Gauss-Bonnet Braneworlds}
The Gauss-Bonnet term with coupling constant $\beta$ is written as
follows
$$L_{GB}=R^{(5)2}-4R_{ab}^{(5)}R^{(5)ab}+R_{abcd}^{(5)}R^{(5)abcd}$$
where $R^{(5)}$ is the curvature scalar of the 5-dimensional bulk
spacetime. These corrections have origin on stringy effects and the
most general action should involve both Gauss-Bonnet and the
Einstein-Hilbert term in 5D theory. The GB term is present only in
the bulk action
\begin{equation}
S_{bulk}=\frac{1}{2\kappa_{5}^{2}}\int d^{5}x\sqrt{-g^{(5)}}\Bigg[
R^{(5)}-2\Lambda_{5}+\beta\Big(R^{(5)2}-4R_{ab}^{(5)}R^{(5)ab}+R_{abcd}^{(5)}R^{(5)abcd}\Big)\Bigg],
\end{equation}
where $\beta$ is the Gauss-Bonnet coupling which can be positive or
negative in the classical GB theory. If $\beta$ is negative, it has
been seen in [42] that this braneworld model leads to antigravity or
tachyon modes on the brane. However, in the presence of a bulk
scalar field, these effects are not present even with negative
$\beta$.\\
The Friedmann equation in the presence of Gauss-Bonnet effects is as
follows [43]
\begin{equation}
H^{2}=\frac{C_{+}+C_{-}-2}{8 \beta}-\frac{K}{a^{2}},
\end{equation}
where
\begin{equation}
C_{\pm}=\Bigg[\sqrt{\Big(1+\frac{4}{3}\beta
\Lambda_{5}+8\beta\frac{\Upsilon}{a^{4}}\Big)^{\frac{3}{2}}+\frac{\beta
\kappa_{5}^{4}(\rho+\lambda)^{2}}{2}}\pm\kappa_{5}^{4}(\rho+\lambda)\sqrt{\frac{\beta}{2}}\Bigg]^{\frac{2}{3}}.
\end{equation}
This equation is a cubic equation with three possible roots. For
$\rho>0$ there is only one real root.\\
The behavior of GB model at high and low energies are as follows
[35]
\begin{equation}
\hspace{1.5 cm}H\gg\alpha^{-\frac{1}{2}}\rightarrow
H^{2}\propto\rho^{\frac{2}{3}},\hspace{2 cm}high\,\, energy\,\,
limit
\end{equation}
and
\begin{equation}
\hspace{1.1 cm}H\ll\alpha^{-\frac{1}{2}}\rightarrow
H^{2}\propto\rho^{2}\hspace{2 cm}low\,\, energy\,\, limit.
\end{equation}
\section{Gauss-Bonnet Induced Gravity with Non-Minimally Coupled Scalar Field on the Brane }
As we have explained, Gauss-Bonnet effect is a high energy stringy
effect. On the other hand, non-minimal coupling of scalar field and
induced gravity on the brane is forced upon us from several
compelling reasons. Some of these reasons have their origin on pure
quantum field theoretical considerations[14]. Then it is natural to
incorporate both Gauss-Bonnet and non-minimal coupling effects to
have a more reliable framework for treating cosmological dynamics.
The action of the GBIG (Gauss-Bonnet term in the bulk and the
Induced Gravity term on the brane) scenario in the presence of a
non-minimally coupled scalar field on the brane can be written as
follows

$$S=\frac{1}{2\kappa_{5}^{2}}\int d^{5}x\sqrt{-g^{(5)}}\Bigg[
R^{(5)}-2\Lambda_{5}+\beta\Big(R^{(5)2}-4R_{ab}^{(5)}R^{(5)ab}+R_{abcd}^{(5)}R^{(5)abcd}\Big)\Bigg]$$
\begin{equation}
+\Bigg[\frac{r}{2\kappa_{5}^{2}}\int
d^{4}x\sqrt{-g}\bigg(\alpha(\phi) R -2\kappa_{4}^{2} g^{\mu\nu}
\nabla_{\mu}\phi\nabla_{\nu}\phi -4\kappa_{4}^{2}V(\phi) -4
\kappa_{4}^{2}\lambda\bigg)\Bigg]_{y=0},
\end{equation}
where $\beta$ and $r$ are the GB coupling constant and IG cross-over
scale respectively. The relation for energy conservation on the
brane is as follows
\begin{equation}
\dot{\rho}+\dot{\rho}_{\phi}+3H(1+\omega)(\rho+\rho_{\phi})=6\alpha'\dot\phi
\Big(H^2+\frac{K}{a^2}\Big).
\end{equation}
where $\omega=\frac{p+p_{\phi}}{\rho+\rho_{\phi}}$\, with $p$ and
$\rho$ pressure and density of ordinary matter. Since for ordinary
matter, $\dot{\rho}+3H(\rho+P)=0$, the non-minimal coupling of the
scalar field and induced curvature on the brane leads to the
non-conservation of the scalar field effective energy
density[41].\\
The cosmological dynamics of the model is given by the following
generalized Friedmann equation
\begin{equation}
\Bigg[1+\frac{8}{3}\beta
\Big(H^{2}+\frac{\Psi}{2}+\frac{K}{a^2}\Big)\Bigg]^{2}\Big(H^2-\Psi+\frac{K}{a^2}\Big)=\Bigg[r
\alpha(\phi)H^{2}+r
\alpha(\phi)\frac{K}{a^2}-\frac{\kappa_{5}^{2}}{6}(\rho+\rho_{\phi}+\lambda)
\Bigg]^{2}.
\end{equation}
This equation describes the cosmological evolution on the brane with
tension and a non-minimally coupled scalar field on the brane. The
bulk contains a black hole mass and a cosmological constant. $\Psi$
is defined as follows
\begin{equation}
\Psi+2 \beta
\Psi^{2}=\frac{\Lambda_{5}}{6}+\frac{\Upsilon}{a^{4}}.
\end{equation}
If $\beta=0$, the model reduces to DGP model, while for $r=0$ we
recover the Gauss-Bonnet model. Here we restrict our study to the
case where bulk black hole mass vanishes, ($\Upsilon=0$) and
therefore $\Psi+2 \beta \Psi^{2}=\frac{\Lambda_{5}}{6}$. The bulk
cosmological constant in the presence of GB term is given by
$\Lambda_{5}=-\frac{6}{l^{2}}+\frac{12\beta}{l^{4}}$, where $l$ is
the bulk curvature. For a spatially flat brane ($K=0$), the
Friedmann equation is given by
\begin{equation}
\Bigg[1+\frac{8}{3}\beta
\Big(H^{2}+\frac{\Psi}{2}\Big)\Bigg]^{2}\Big(H^2-\Psi\Big)=\Bigg[r
\alpha(\phi)H^{2}-\frac{\kappa_{5}^{2}}{6}(\rho+\rho_{\phi}+\lambda)
\Bigg]^{2}.
\end{equation}
We define the following dimensionless quantities
\begin{equation}
\gamma=\frac{8\beta}{3r^{2}},\hspace{1cm}\chi=\frac{r^{2}}{l^{2}},\hspace{1cm}\psi=\Psi
r^{2},
\end{equation}
where the dimensionless Friedmann equation takes the following form
\begin{equation}
\Bigg[1+\gamma
\Big(h^{2}+\frac{\psi}{2}\Big)\Bigg]^{2}\Big(h^2-\psi\Big)=\Bigg[
\alpha(\phi)h^{2}-\Big(\mu+ \mu'+\sigma
-2\frac{d\alpha(\phi)}{d\tau}h \kappa_{4}^{2}\Big) \Bigg]^{2}.
\end{equation}
To find cosmological dynamics of our model, we should solve this
equation in an appropriate parameter space. In which follows, we
consider Minkowski and AdS bulk and investigate their cosmological
consequences.
\subsection{Minkowski Bulk($\psi=0$) with Tensionless Brane($\sigma=0$)}
In this case, the non-minimal GBIG Friedmann equation takes the
following form
\begin{equation}
\Big(1+\gamma h^{2}\Big)^{2} h^{2}=\Bigg[
\alpha(\phi)h^{2}-\Big(\mu+ \mu' -2\frac{d\alpha(\phi)}{d\tau}h
\kappa_{4}^{2}\Big) \Bigg]^{2}.
\end{equation}
It is straightforward to show that in this case
$$\frac{d(\mu+\mu')}{d(h^2)}=-\frac{(1+\gamma h^{2})(3\gamma
h^{2}+1)}{2\Big(\alpha(\phi)h^{2}-(\mu+\mu')+2\frac{d\alpha(\phi)}{d\tau}h
\kappa_{4}^{2}\Big)}\hspace{5cm}$$
\begin{equation}
+\frac{2\Bigg[\alpha(\phi)^{2}h^{2}+3\alpha(\phi)\frac{d\alpha(\phi)}{d\tau}h
\kappa_{4}^{2}-(\mu+\mu')\Big(\alpha(\phi)+\frac{1}{h}\frac{d\alpha(\phi)}{d\tau}
\kappa_{4}^{2}\Big)+2(\frac{d\alpha(\phi)}{d\tau}
\kappa_{4}^{2})^{2}\Bigg]}{2\Big(\alpha(\phi)h^{2}-(\mu+\mu')+2\frac{d\alpha(\phi)}{d\tau}h
\kappa_{4}^{2}\Big)}.
\end{equation}
The initial Hubble rate and density are given by substituting the
result of $\frac{d(\mu+\mu')}{d(h^2)}=0$ in to equation (24). This
leads us to the following relations
\begin{equation}
h_{i}=\frac{\alpha(\phi)+\sqrt{\alpha(\phi)^{2}-3\gamma+6\gamma\frac{d\alpha(\phi)}{d\tau}
\kappa_{4}^{2}}}{3\gamma},
\end{equation}
\begin{equation}
(\mu+\mu')_{i}=\frac{2\alpha(\phi)^{3}-9\alpha(\phi)\gamma+18\alpha(\phi)\frac{d\alpha(\phi)}{d\tau}
\kappa_{4}^{2}\gamma+2\Big(\alpha(\phi)^{2}-3\gamma+6\gamma\frac{d\alpha(\phi)}{d\tau}
\kappa_{4}^{2}\Big)^{\frac{3}{2}}}{27\gamma^{2}}.\hspace{7 cm}
\end{equation}

Before proceeding further, we should stress on two important points
here: firstly, the presence of GB term removes the big bang
singularity in this setup, and the universe starts with an initial
finite density. Gauss-Bonnet effect is essentially a string-inspired
effect in the bulk which its combination with pure DGP scenario
leads to a finite big bang proposal on the brane. A consequence of
string inspired field theories is the existence of minimal
observable length of the order of Planck length[44-46]. One cannot
probe distances smaller than this fundamental length. In fact a
string cannot live on the scale smaller than its length. This
feature leads us to generalize the standard Heisnberg uncertainty
relation to incorporate this Planck scale effect[47,48]. The
existence of this minimal observable length essentially removes
spacetime singularity and acts as a UV cutoff of the corresponding
field theory( see for instance [49] which discusses inflation with
minimum length cutoff. See also [50]). So, in principle existence of
a finite density big bang is supported at least in this
viewpoint[51], see also [52]. Secondly, non-minimal coupling of
scalar field with induced gravity on the brane controls the value of
the initial density. This is not the only importance of non-minimal
coupling of scalar field and induced gravity. In fact non-minimal
coupling provides a mechanism for generating spontaneous symmetry
breaking at Planck scale on the brane[53]. In this respect and based
on the arguments presented at the introduction on the importance of
the non-minimal coupling, non-minimal coupling of scalar field and
induced gravity on the brane itself is a high energy correction of
the theory and it is natural to expect that this effect couples with
stringy effects in Planck scale. In fact in this setup, we encounter
a smoother behavior due to Gauss-Bonnet term ( a finite density big
bang) and the late-time effects of non-minimally coupled scalar
field component. These effects together provide a more reliable
cosmological scenario.

The behavior of $h$ with respect to $\mu+\mu'$ is shown in figure
$5$. This figure shows also that the GBIG1 and GBIG2 branches have
self-acceleration for some positive values of non-minimal coupling
in the same manner as the DGP(+) branch, whereas GBIG3 branch
similar to DGP(-) has no self-acceleration. This is similar to pure
DGP or GB model alone where there is a big bang singularity. The
self-accelerating GBIG2 branch is not a physical solution since it
is accelerating throughout its evolution. For negative values of
non-minimal coupling, the GBIG3 and GBIG2 have self-acceleration
while GBIG1 has not such a property. Now it is easy to show that
$$(\mu+\mu')\rightarrow\infty :\hspace{1cm}\gamma\rightarrow0 \hspace{2.5cm}$$
$$(\mu+\mu')\rightarrow 0 :\hspace{1cm}\gamma\rightarrow \frac{\alpha(\phi)^{2}}{4(1-2\frac{d \alpha}{d
\tau}\kappa_{4}^{2})}.$$ In the minimal case, the maximum value of
$\gamma$ leads to a minimum value of $h_{i}$. Here, in the presence
of non-minimal coupling of scalar field and induced gravity with
$\gamma_{max}= \frac{\alpha(\phi)^{2}}{4(1-2\frac{d \alpha}{d
\tau}\kappa_{4}^{2})}$\,, we cannot conclude that this leads to a
minimum value for $h$. Using equation (26), the value of $h_{i}$ for
$\gamma_{max}$ is given as follows
\begin{equation}
h_{i}=\frac{2(1-2\frac{d \alpha}{d
\tau}\kappa_{4}^{2})}{\alpha(\phi)}.
\end{equation}
When $\gamma=\gamma_{max}$ and $(\mu+\mu')= 0$, there is a vacuum
brane with de Sitter expansion. The $h$ asymptotic value is
obtained when $\mu+\mu'\rightarrow 0$ in equation (24)
\begin{equation}
h_{\infty}^{6}+\frac{(2\gamma-\alpha^{2})}{\gamma^{2}}h_{\infty}^{4}-\frac{(4\alpha
\frac{d \alpha}{d
\tau}\kappa_{4}^{2})}{\gamma^{2}}h_{\infty}^{3}+\frac{\Big[1
-4(\frac{d \alpha}{d
\tau}\kappa_{4}^{2})^{2}\Big]}{\gamma^{2}}h_{\infty}^{2}=0.
\end{equation}
This equation has four non-zero roots that two of them are
negative and unacceptable. When $\gamma\rightarrow 0$, we should
have the non-minimal DGP model.
\begin{figure}[htp]
\begin{center}
\includegraphics{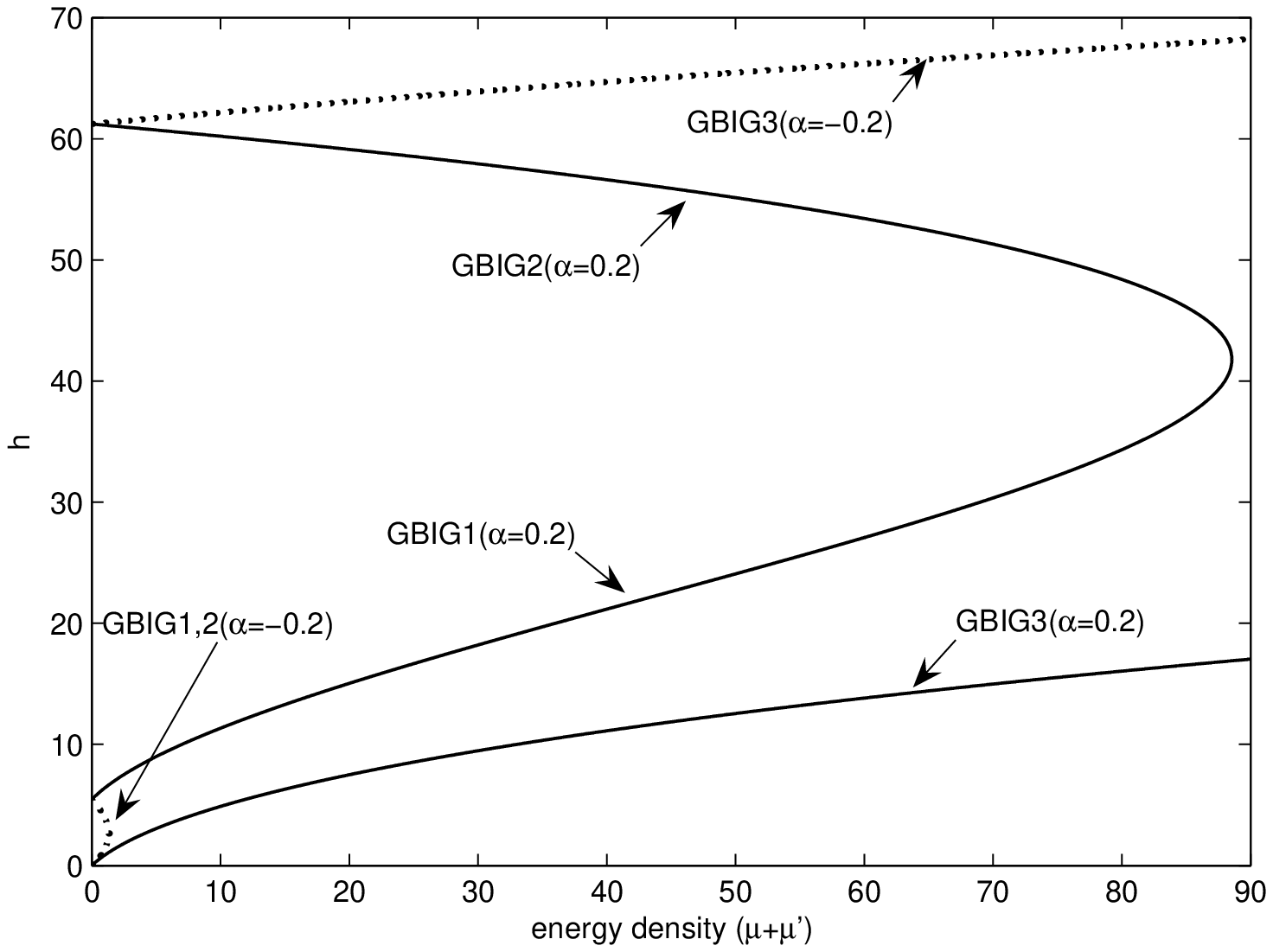} \vspace{5cm}\includegraphics{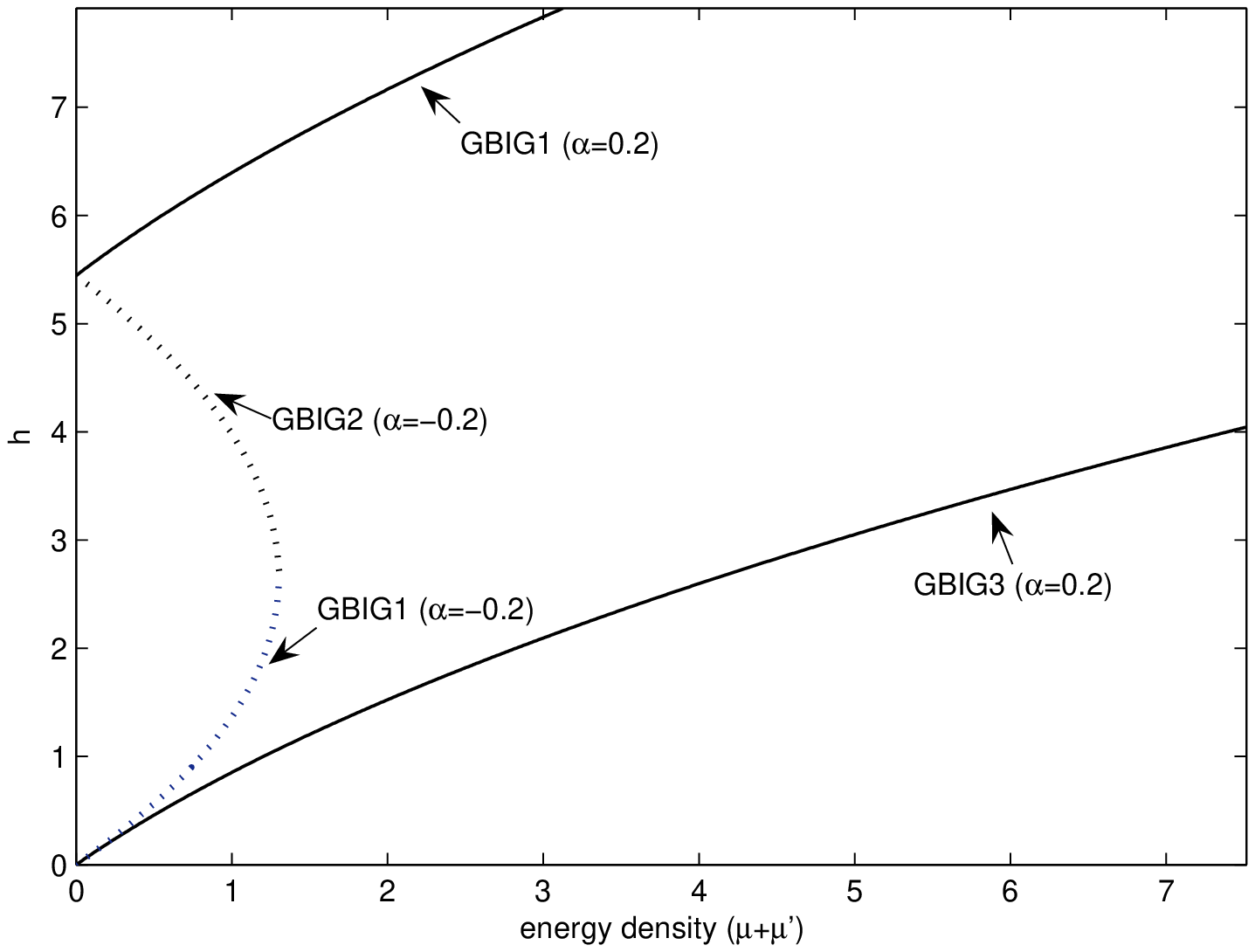}
\end{center}
 \caption{\small { Solutions of the Friedmann equation for a tensionless brane in a Minkowski
 bulk. For clarification, we have plotted the figure in two different scales to highlight intermediate
 points. Also we have set $\gamma=0.003$ and $\alpha(\phi)=0.2, -0.2$.}}
 \end{figure}\\
From equation (24) one can deduce
\begin{equation}
(\mu+\mu')=\alpha(\phi)h^{2}-h(\gamma h^{2}+1)+2\frac{d \alpha}{d
\tau}\kappa_{4}^{2}h,
\end{equation}
where $h_{\infty}\leq h<h_{i}$. Since $(\mu+\mu')_{\infty}$=0, from
equation (30) it follows that
\begin{equation}
\gamma=\frac{\alpha(\phi)h_{\infty}-1+2\frac{d \alpha}{d
\tau}\kappa_{4}^{2}}{h_{\infty}^{2}}.
\end{equation}
By expanding $\mu+\mu'$ to first order in $h^{2}-h_{\infty}^{2}$,
we find
\begin{equation}
h^{2}=h_{\infty}^{2}+\frac{2\Big(\alpha(\phi)h_{\infty}^{2}+2\frac{d
\alpha}{d
\tau}\kappa_{4}^{2}h_{\infty}\Big)}{\alpha(\phi)h_{\infty}\Big(2-6\frac{d
\alpha}{d
\tau}\kappa_{4}^{2}-\alpha(\phi)h_{\infty}\Big)-8(\frac{d
\alpha}{d \tau}\kappa_{4}^{2})^{2}+4\frac{d \alpha}{d
\tau}\kappa_{4}^{2}}(\mu+\mu').
\end{equation}
In comparison with equation (5), we find the following effective
4-dimensional Newton's constant
\begin{equation}
G=\Bigg[\frac{\Big(\alpha(\phi)h_{\infty}^{2}+2\frac{d \alpha}{d
\tau}\kappa_{4}^{2}h_{\infty}\Big)}{\alpha(\phi)h_{\infty}\Big(2-6\frac{d
\alpha}{d
\tau}\kappa_{4}^{2}-\alpha(\phi)h_{\infty}\Big)-8(\frac{d
\alpha}{d \tau}\kappa_{4}^{2})^{2}+4\frac{d \alpha}{d
\tau}\kappa_{4}^{2}}\Bigg]\frac{\alpha(\phi)G_{5}}{r},
\end{equation}
where $G_{5}=\frac{\kappa_{5}^{2}}{8\pi}$ and
$G=\frac{\kappa_{4}^{2}}{8\pi}$ are five and four dimensional
gravitational constant respectively. From equation (33) we obtain
a relation between $M_{5}^{3}$ and $M_{p}^{2}$ as follows
\begin{equation}
M_{5}^{3}\simeq \Bigg[\frac{\Big(\alpha(\phi)r^{2}H_{0}^{2}+2\frac{d
\alpha}{d\tau}\kappa_{4}^{2}rH_{0}\Big)}{\alpha(\phi)rH_{0}\Big(2-6\frac{d
\alpha}{d \tau}\kappa_{4}^{2}-\alpha(\phi)rH_{0}\Big)-8(\frac{d
\alpha}{d \tau}\kappa_{4}^{2})^{2}+4\frac{d \alpha}{d
\tau}\kappa_{4}^{2}}\Bigg]\frac{\alpha(\phi)M_{p}^{2}}{r}.
\end{equation}
Here $H_{\infty}\sim H_{0}$, and we see the important role played by
non-minimal coupling in this setup. In principle, one can fine tune
the value of the non-minimal coupling such that fundamental scale of
the bulk be reduced to values in the range accessible for next
generation of accelerators. To compare with DGP(+) limit, when
$\mu+\mu'\rightarrow0$, that is at late-time, we have
$rH_{0}\longrightarrow\frac{1}{\alpha(\phi)}-2\frac{\kappa_{4}^{2}}{\alpha(\phi)}\frac{d
\alpha}{d \tau}$, therefore this equation in this limit reduces to
$M_{5}^{3}\simeq \frac{M_{p}^{2}}{r}$. This is an interesting result
since there is no effect of non-minimal coupling in the non-minimal
DGP(+) limit at late-time and the relation between $M_{5}^{3}$ and
$M_{p}^{2}$ is the same as minimal case. This is not surprising
since essentially a part of motivation for inclusion of non-minimal
coupling has its origin on the quantum field theoretical
considerations( the renormalizability of quantum field theory in
curved background and quantum corrections to the scalar field
theory). In this view point, non-minimal coupling shows its
importance mainly in the high energy UV sector of the theory while
apparently DGP(+) gives IR sector of the theory free of stringy and
strong quantum field theoretical effects. We should stress that
non-minimal coupling of scalar field and induced gravity on DGP
brane modifies cross over scale[39]. The above argument is
restricted to the limit $\mu+\mu'\rightarrow0$ which is related to
the late time stage of evolution. In equation (34), when
$\alpha(\phi)$ attains different values, $M_{5}$ increases or
decreases relative to its value in DGP limit. When
$rH_{0}\rightarrow\frac{2(1-2\frac{d
\alpha}{d\tau}\kappa_{4}^{2})}{\alpha(\phi)}$, $M_{5}$ increases. We
should stress that these results are sensitive to the sign of the
non-minimal coupling explicitly.

\subsection{Minkowski Bulk ($\psi=0$) with Brane Tension($\sigma\neq0$)}

In this case the Friedmann equation is as follows
\begin{equation}
(1+\gamma h^{2})^{2}h^2=\Bigg[ \alpha(\phi)h^{2}-\Big(\mu+
\mu'+\sigma -2\frac{d\alpha(\phi)}{d\tau}h \kappa_{4}^{2}\Big)
\Bigg]^{2}.
\end{equation}
The effect of brane tension is similar to considering a cosmological
constant on the brane. In this case, there are three possible
solutions (GBIG\,1-3). To find initial and final Hubble rates and
density, we set $\frac{d(\mu+\mu')}{d(h^{2})}=0$. The initial and
final Hubble rates denoted by $h_{i}$ and $h_{e}$ respectively, are
given by
\begin{equation}
h_{i,\,e}=\frac{\alpha(\phi)\pm\sqrt{\alpha(\phi)^{2}-3\gamma+6\gamma\frac{d\alpha(\phi)}{d\tau}
\kappa_{4}^{2}}}{3\gamma},
\end{equation}
and the initial and final density are calculated as follows
\begin{equation}
(\mu+\mu')_{i,\,e}=\frac{2\alpha(\phi)^{3}-9\alpha(\phi)\gamma+18\alpha(\phi)\frac{d\alpha(\phi)}{d\tau}
\kappa_{4}^{2}\gamma\pm2\Big(\alpha(\phi)^{2}-3\gamma+6\gamma\frac{d\alpha(\phi)}{d\tau}
\kappa_{4}^{2}\Big)^{\frac{3}{2}}}{27\gamma^{2}}-\sigma,\hspace{7
cm}
\end{equation}
where the plus sign is for initial state and minus sign shows the
final state. These points have $h'=0$.  The point $h=0$ is the place
which GBIG3 loiters, that it is given by
\begin{equation}
(\mu+\mu')_{l}=-\sigma.
\end{equation}
and $h_{\infty}$ is given by
$$h_{\infty}^{6}+\frac{(2\gamma-\alpha^{2})}{\gamma^{2}}h_{\infty}^{4}-\frac{(4\alpha
\frac{d \alpha}{d
\tau}\kappa_{4}^{2})}{\gamma^{2}}h_{\infty}^{3}+\frac{\Big[1+2\alpha\sigma
-4(\frac{d \alpha}{d
\tau}\kappa_{4}^{2})^{2}\Big]}{\gamma^{2}}h_{\infty}^{2}$$
\begin{equation}
+\frac{(4\sigma \frac{d \alpha}{d
\tau}\kappa_{4}^{2})}{\gamma^{2}}h_{\infty}-\frac{\sigma^{2}}
{\gamma^{2}}=0.
\end{equation}
to obtain this relation we have set $\mu+\mu'=0$ in equation (35).\\
The relation between $h$ and $\mu+\mu'$ is shown in figure $6$ for a
negative tension brane. In this figure for positive non-minimal
coupling, the GBIG1 and GBIG2 branches start with an initial Hubble
rate and density $\Big(h_{i},(\mu+\mu')_{i}\Big)$ whereas the GBIG3
branch doesn't remove the big bang singularity. GBIG1 and GBIG3
terminate at a finite Hubble rate and density
$\Big(h_{e},(\mu+\mu')_{e}\Big)$, whereas GBIG2 terminates in a
vacuum de Sitter state. On the other hand, for negative non-minimal
coupling, the GBIG2 branch has a big bang singularity and this is a
self-accelerating solution. GBIG1 and GBIG3 start without big bang
singularity $\Big(h_{i},(\mu+\mu')_{i}\Big)$ and both of them have
self-acceleration but GBIG3 throughout evolution loiters and then
evolves. A universe which undergoes a period of loitering is an
attractive alternative to standard cosmologies. Generally a
loitering universe is an expanding Friedmann universe that undergoes
a phase of slow expansion with redshift of  $(z \sim 3 - 5)$. It is
believed that the large scale structure of the universe is formed
during this semi-static phase[54].
\begin{figure}[htp]
\begin{center}
\includegraphics{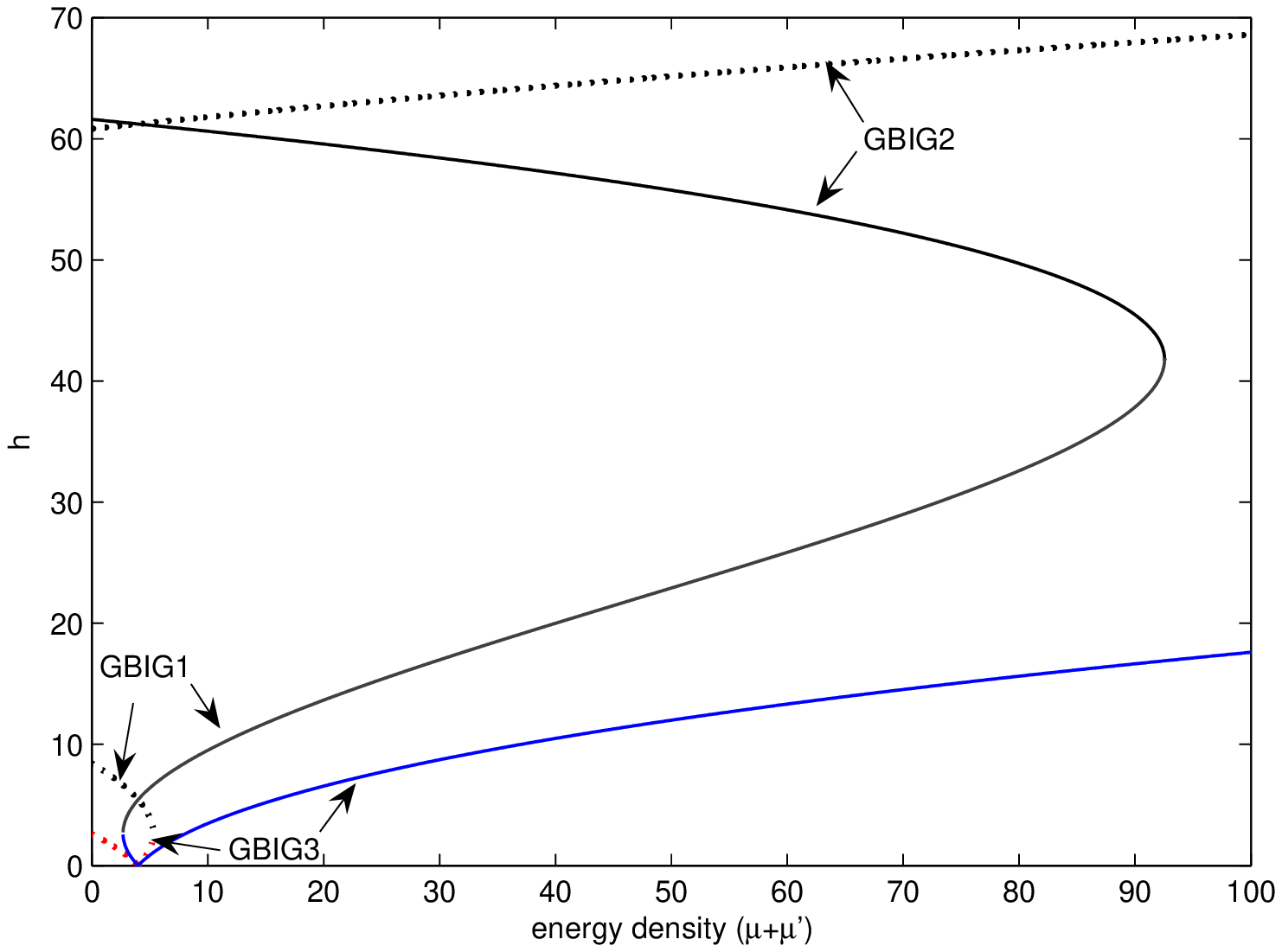} \vspace{5cm}\includegraphics{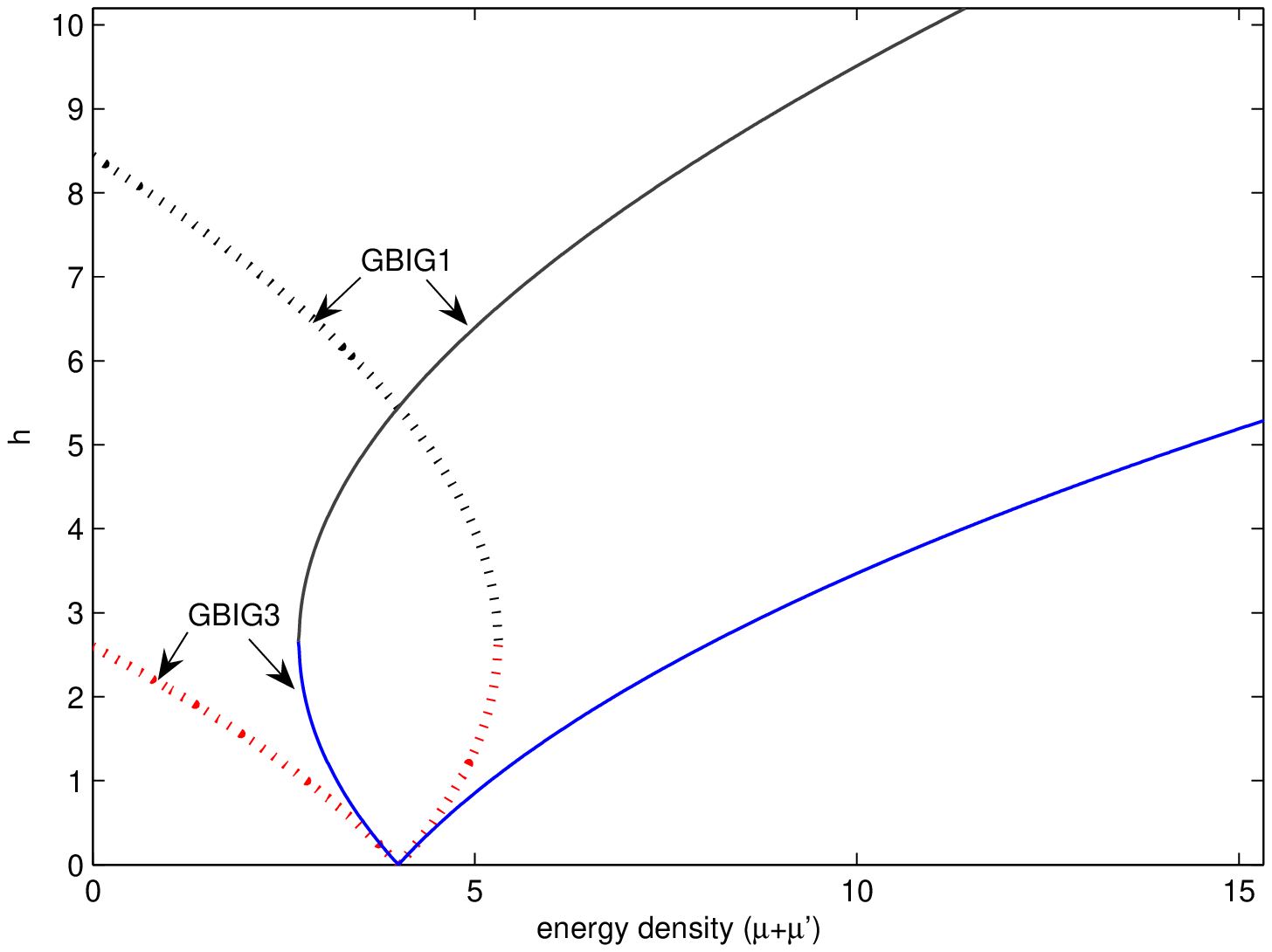}
\end{center}
 \caption{\small { Solutions of the Friedmann equation with a negative tension brane in a Minkowski
 bulk. We have assumed
 $\gamma=0.003$, $\sigma=-4$, $\psi=0$, $\alpha(\phi)=0.2$ and $-0.2$ for solid and dot curves
 respectively. The figures are plotted with different scale to highlight important features.}}
\end{figure}
\newpage
Figure $7$ is the result for a positive tension brane. Here with
positive non-minimal coupling, the three solutions have
self-acceleration. There are a finite density for GBIG1 and GBIG2
but GBIG3 has big bang singularity. For negative non-minimal
coupling, these solutions are not physically reliable since for all
of them the density is negative.\\
\begin{figure}[htp]
\begin{center}
\includegraphics{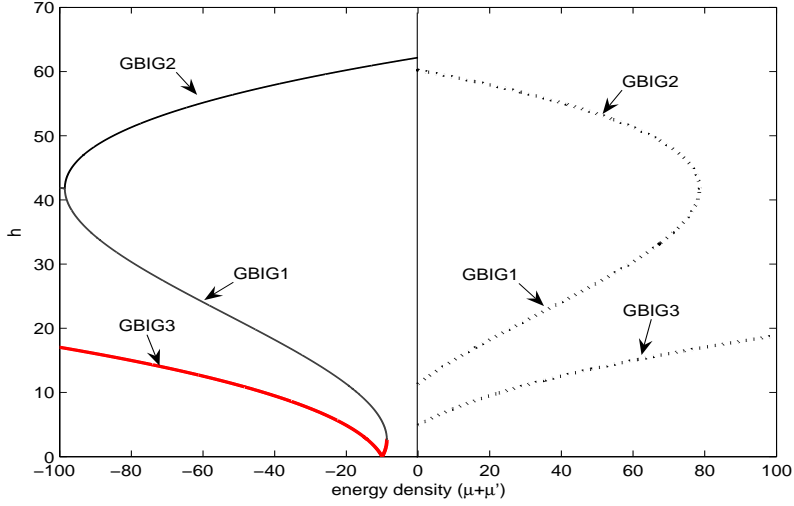}
\end{center}
\vspace{6 cm}
  \caption{\small { Solutions of the Friedmann equation with a positive tension brane in a Minkowski bulk.
 $\gamma=0.003$, $\sigma=10$, $\psi=0$, $\alpha(\phi)=0.2$ and $-0.2$ for dot and solid curves respectively.}}
\end{figure}

Note also that $\sigma_{i,\,e}$ and $\sigma_{l}$ are quantities
for which $\mu_{i,\,e}=0$ and $\mu_{l}=0$ respectively. From
equation (37), $\sigma_{i,\,e}$ are given in terms of $\gamma$ and
$\alpha(\phi)$ by
\begin{equation}
\sigma_{i,\,e}=\frac{2\alpha^{3}(\phi)-9\alpha(\phi)\gamma+18\alpha(\phi)\frac{d\alpha(\phi)}{d\tau}
\kappa_{4}^{2}\gamma\pm2\Big(\alpha^{2}(\phi)-3\gamma+6\gamma\frac{d\alpha(\phi)}{d\tau}
\kappa_{4}^{2}\Big)^{\frac{3}{2}}}{27\gamma^{2}},
\end{equation}
and from equation (38), $\sigma_{l}=0$. In the non-minimal case,
the maximum value of $\gamma$ is
$\gamma_{max}=\frac{\alpha^{2}(\phi)}{3-6\frac{d\alpha(\phi)}{d\tau}
\kappa_{4}^{2}}$. For
$\gamma=\frac{\alpha^{2}(\phi)}{3-6\frac{d\alpha(\phi)}{d\tau}
\kappa_{4}^{2}}$, GBIG1 branch disappears, since $h_{i}=h_{e}$ at
$\gamma_{max}$. Actually the point $h_{i}=h_{e}$ is now a point of
inflection. The requirement to have real value for the square root
in equation (36) and (37) leads us to the following relation
\begin{equation}
\gamma\leq\frac{\alpha^{2}(\phi)}{3-6\frac{d\alpha(\phi)}{d\tau}
\kappa_{4}^{2}}.
\end{equation}
Based on this relation, the range of variation of $\gamma$ depends
on the non-minimal coupling coefficient directly. The latest
observational constraints on the values of $\gamma$ are listed in
table $1$ [55]. In this framework, we can constraint non-minimal
coupling of scalar field and induced Ricci scalar using
observational data. Considering a conformal coupling of the scalar
field and induced gravity on the brane defined as
$\alpha(\phi)=(1-\xi \phi^{2})$, we can constraint $\xi$ based on
the constraints imposed on $\gamma$ presented in table $1$. A
detailed study of constraints on non-minimal coupling of scalar
field and gravity based on various observational data and
theoretical techniques are summarized in reference [38]( see also
[56-64] for more details). For a time-independent scalar field,
$\alpha(\phi)$ will be time-independent also. Using the result of
SNIa+LSS+H(z)( the third line of table $1$) for $\gamma$ and
relation (41), we obtain $\alpha\geq0.01$ and $\alpha\leq -0.01$.
Since $\alpha(\phi)=(1-\xi \phi^{2})$, we find
$$-\sqrt{\frac{0.99}{\xi}}\leq\phi\leq\sqrt{\frac{0.99}{\xi}},$$ for $\alpha\geq0.01$ and
$$\phi\geq\sqrt{\frac{1.01}{\xi}}\,\,\,\,\,\,\, and\,\,\,\,\,\,\,
\phi\leq-\sqrt{\frac{1.01}{\xi}},$$ for $\alpha\leq -0.01$
respectively. The values of $\xi$ are constraint to be in the range
of $\xi\leq 0.99\phi^{-2}$ and $\xi\geq 1.01\phi^{-2}$. As we see
these constraints are dependent on the scalar field and this is
reasonable since dynamics of scalar field essentially affects its
coupling to gravity. In this viewpoint, variation of gravitational
coupling as a field in scalar-tensor gravity can be attributed to
variation of non-minimal coupling.
\begin{table*}
\begin{center}
\caption{The values of $\gamma$ in the different test.} \label{t1}
\begin{tabular}{|c|c|c|c|c|c|c|c|}
  \hline
  \hline
    Test& $\gamma$  \\
  \hline
   SNIa &$0.0278_{-0.0278}^{+0.0033}$  \\
   SNIa+LSS& $0.000_{-0.000}^{+0.005}$ \\
   SNIa+LSS+H(z)& $0.000_{-0.000}^{+0.003}$   \\
  \hline
  \hline
\end{tabular}
\end{center}
\end{table*}

\subsection{AdS Bulk ($\psi\neq0$) with Brane Tension($\sigma\neq0$)}
For $\psi\neq0$, the bulk is AdS since $\Lambda_{5}\neq0$. We
should solve equation (23) for this case. The condition $h=0$
gives two solutions
\begin{equation}
\mu_{b,c}=\mp\sqrt{-\psi}(1+\frac{\psi}{2})-\sigma,
\end{equation}
where $\mu_{c}$ is the density at the point that GBIG3 collapses
(corresponding to plus sign), while $\mu_{b}$ is the density of the
new bouncing point for GBIG4 (corresponding to minus sign). In the
previous subsection (Minkowski bulk) there is a loitering point.
Here this point separates into bouncing and collapsing points. It is
interesting that when $\psi=-5$, the GBIG4 branch disappears.
Although the exact value of $\psi$ for disappearance of this branch
is not important and depends on the choice of parameters, the fact
that in principle this branch can be disappeared in a suitable
parameter space is an important result. To obtain turning points of
the branches, we calculate $\frac{d(\mu+\mu')}{d(h^2)}$ as follows
$$\frac{d(\mu+\mu')}{d(h^2)}=-\frac{\Big[1+\gamma (h^{2}+\frac{\psi}{2})\Big]\Big[3\gamma
(h^{2}-\frac{\psi}{2})+1\Big]}{2\Big(\alpha(\phi)h^{2}-(\mu+\mu'+\sigma)+2\frac{d\alpha(\phi)}{d\tau}h
\kappa_{4}^{2}\Big)}\hspace{5cm}$$
\begin{equation}
+\frac{2\Bigg[\alpha(\phi)^{2}h^{2}+3\alpha(\phi)\frac{d\alpha(\phi)}{d\tau}h
\kappa_{4}^{2}-(\mu+\mu'+\sigma)\Big(\alpha(\phi)+\frac{1}{h}\frac{d\alpha(\phi)}{d\tau}
\kappa_{4}^{2}\Big)+2(\frac{d\alpha(\phi)}{d\tau}
\kappa_{4}^{2})^{2}\Bigg]}{2\Big(\alpha(\phi)h^{2}-(\mu+\mu'+\sigma)+2\frac{d\alpha(\phi)}{d\tau}h
\kappa_{4}^{2}\Big)}.
\end{equation}
By substituting $\frac{d(\mu+\mu')}{d(h^2)}=0$, these points can be
obtained by solving the equation
\begin{equation} \frac{3}{2}\gamma
h^{3}+(\frac{1}{2}-\frac{3}{4}\gamma \psi) h-\alpha
h\sqrt{h^{2}-\psi}-\frac{d \alpha}{d
\tau}\kappa_{4}^{2}\sqrt{h^{2}-\psi}=0.
\end{equation}
There are three roots (which are presented in Appendix A ) two of
which are complex. Using equation (23), $h_{\infty}$ for AdS bulk
satisfies the following equation
$$h_{\infty}^{6}+\frac{(2\gamma-\alpha^{2})}{\gamma^{2}}h_{\infty}^{4}-\frac{(4\alpha
\frac{d \alpha}{d
\tau}\kappa_{4}^{2})}{\gamma^{2}}h_{\infty}^{3}+\frac{\Big[1+2\alpha\sigma-\psi\gamma(1+\frac{3}{4}\psi\gamma)
-4(\frac{d \alpha}{d
\tau}\kappa_{4}^{2})^{2}\Big]}{\gamma^{2}}h_{\infty}^{2}$$
\begin{equation}
+\frac{(4\sigma \frac{d \alpha}{d
\tau}\kappa_{4}^{2})}{\gamma^{2}}h_{\infty}-\frac{\Big[\psi(1+\frac{\psi\gamma}{2})^{2}+\sigma^{2}\Big]}
{\gamma^{2}}=0.
\end{equation}
In comparison with the minimal case [35], the behavior of the
branches are changed considerably. To see these differences, we
obtain numerical solutions of the above equation for different
values of $\psi$. The results of this calculations are shown in
figures $8$, $9$ and $10$ respectively.

\begin{figure}[htp]
\begin{center}
\includegraphics{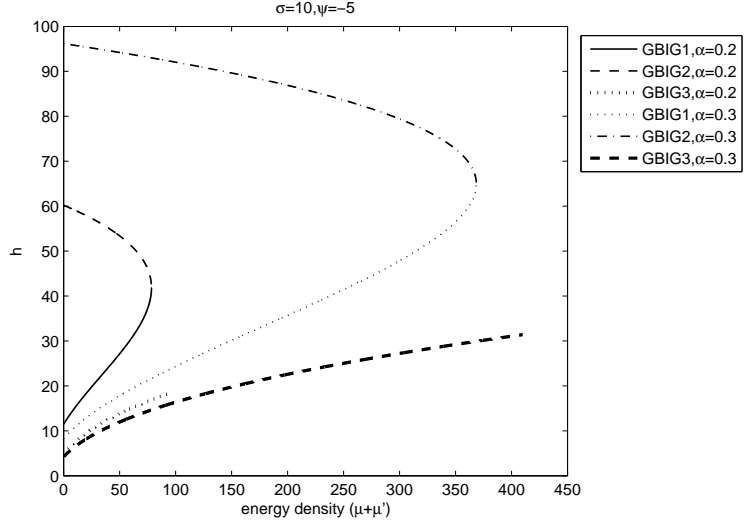}
\end{center}
\vspace{7 cm}
 \caption{\small { The solutions of the Friedmann equation with a positive brane tension in  AdS bulk.
 We have chosen $\gamma=0.003$, $\sigma=10$, $\psi=-5$, $\alpha(\phi)=0.2,\, 0.3$.}}
\end{figure}

\begin{figure}[htp]
\begin{center}
\includegraphics{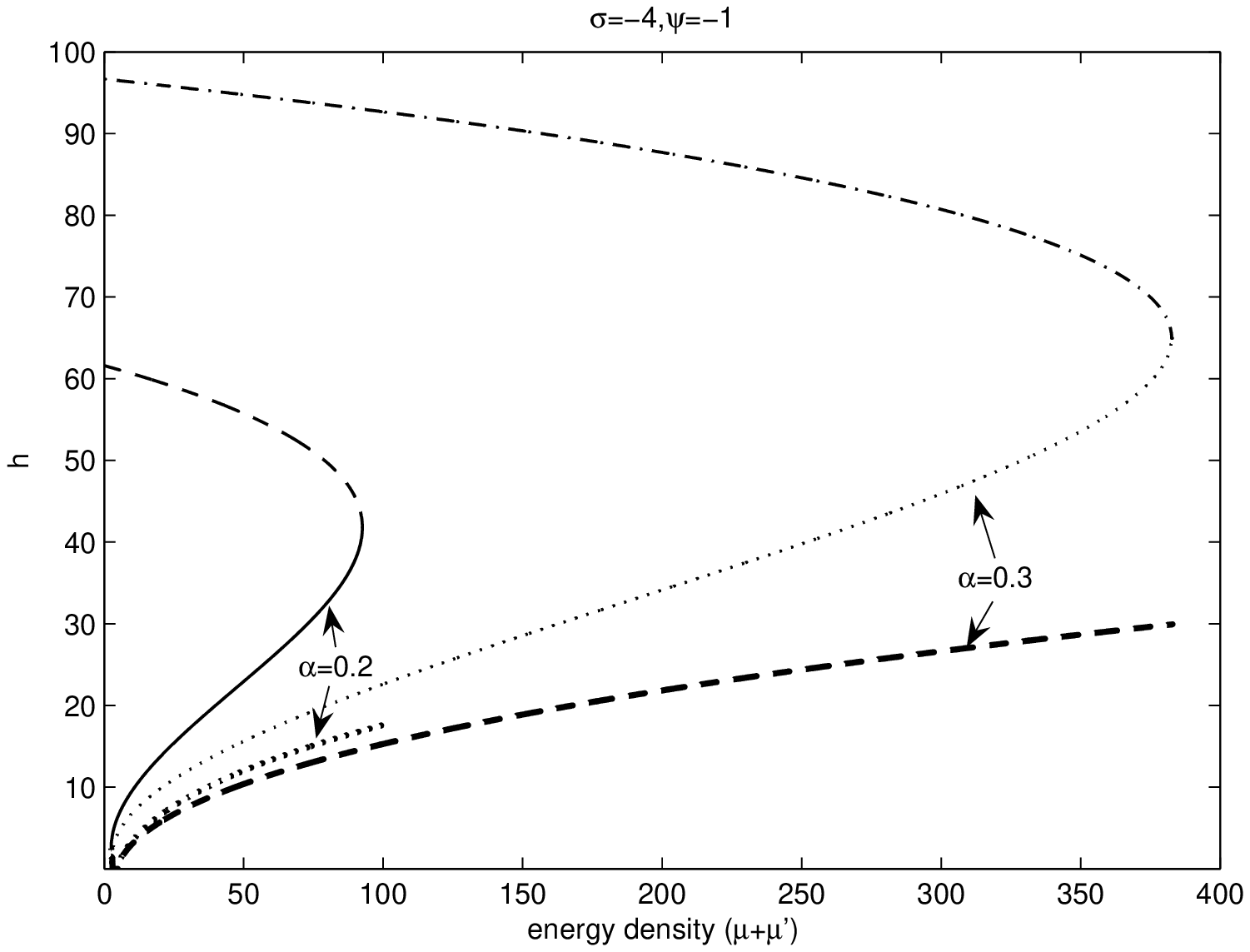} \vspace{5cm}\includegraphics{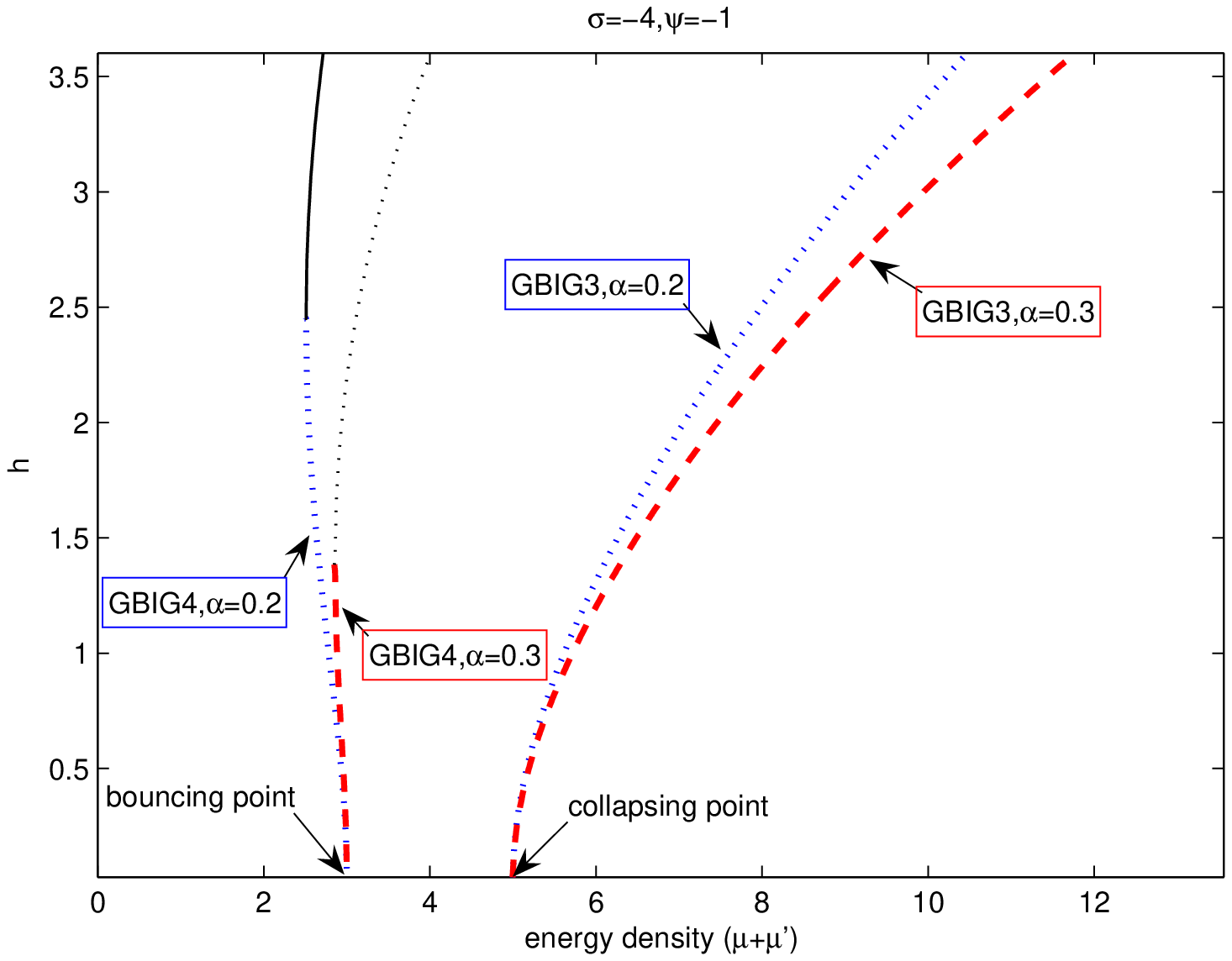}
\end{center}
 \caption{\small { The solutions of the Friedmann equation with a negative brane tension in  AdS bulk and in two different scales.
 We have set $\gamma=0.003$, $\sigma=-4$, $\psi=-1$, $\alpha(\phi)=0.2,\, 0.3$.}}
\end{figure}

\begin{figure}[htp]
\begin{center}
\includegraphics{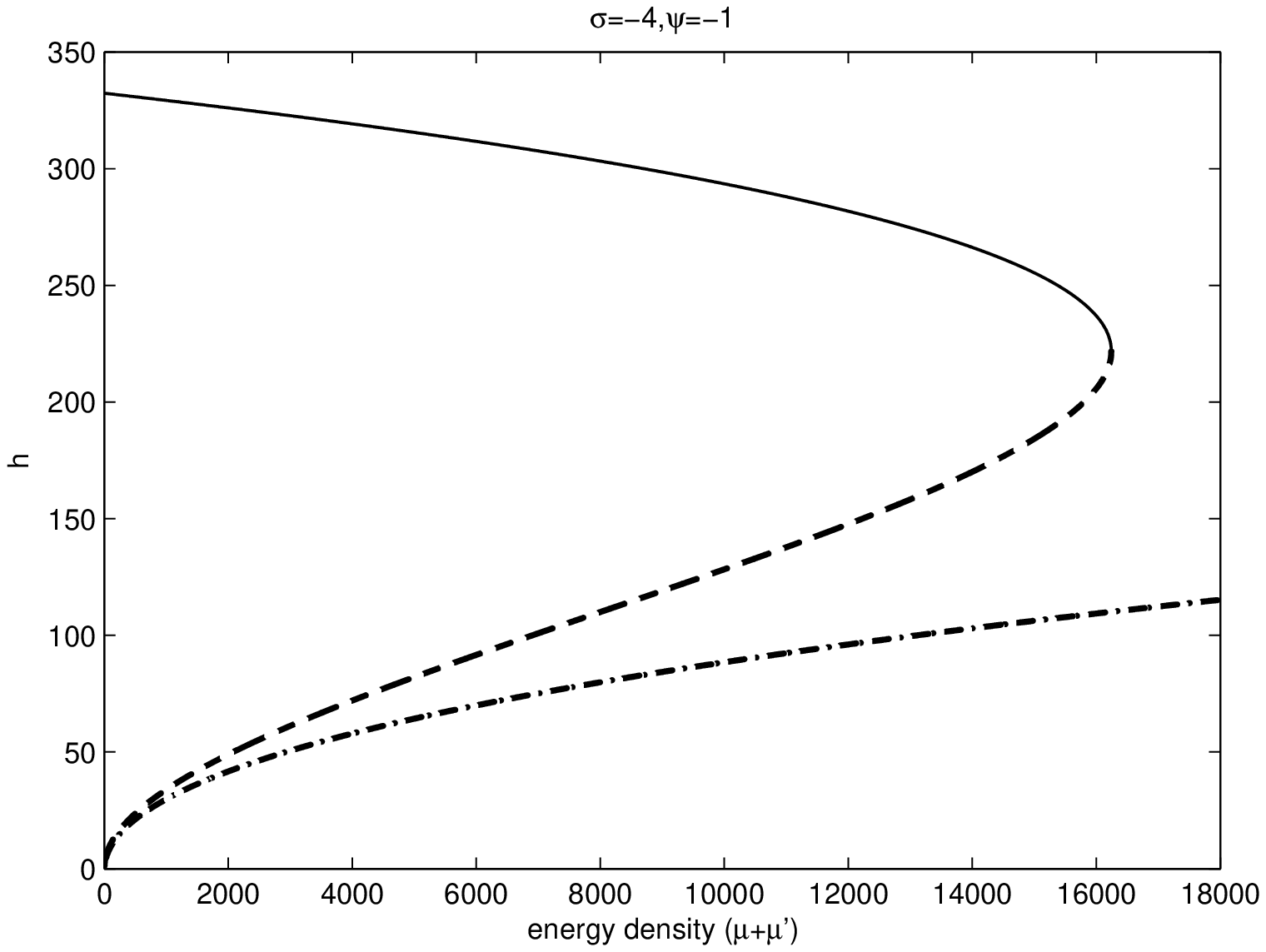} \vspace{5cm}\includegraphics{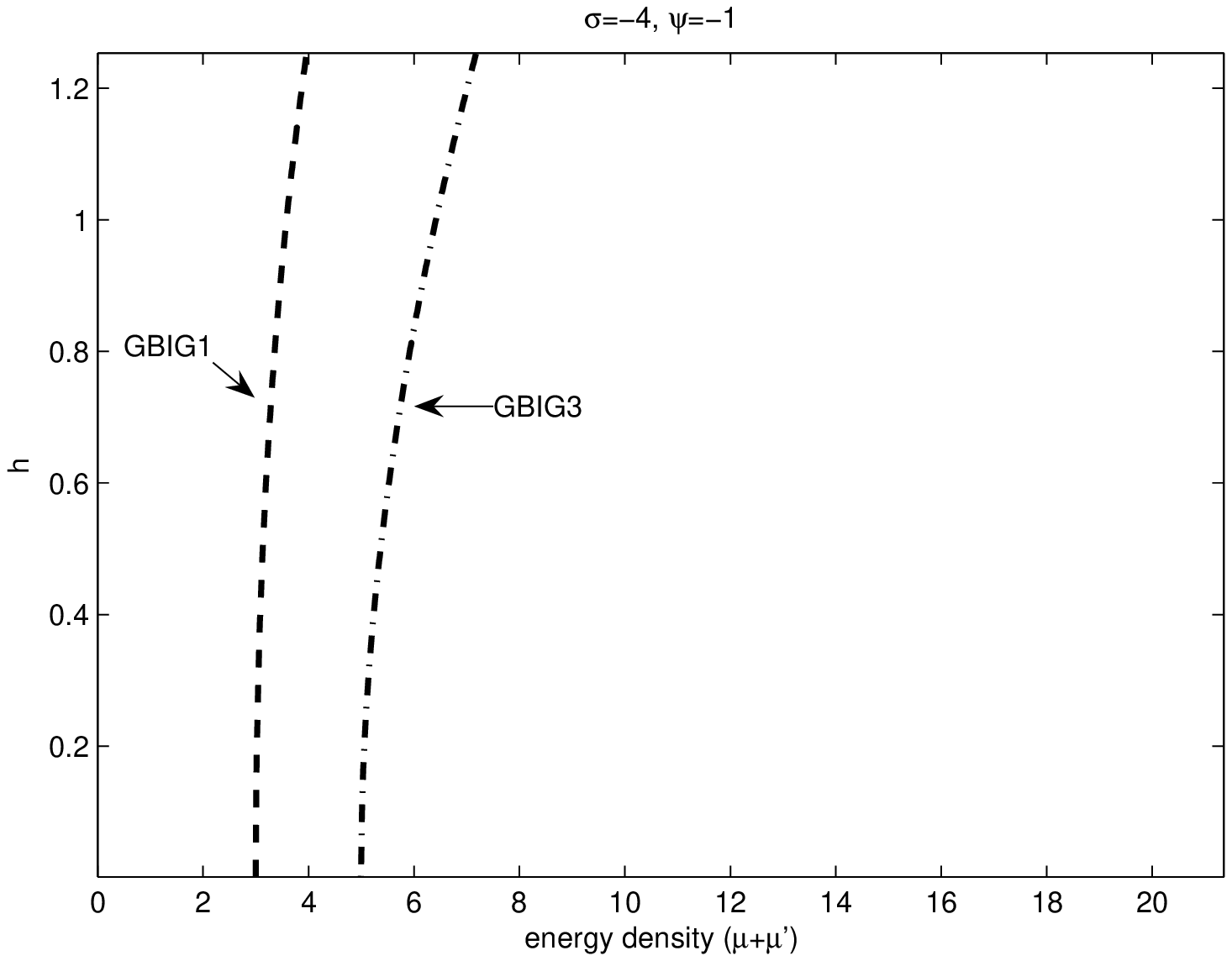}
\end{center}
 \caption{\small { The solutions of the Friedmann equation with a negative brane tension in AdS bulk and in two different scales.
 $\gamma=0.003$, $\sigma=-4$, $\psi=-1$, $\alpha(\phi)=1$ i.e. minimal case.}}
\end{figure}
As these figures show, $\alpha(\phi)$ as non-minimal coupling of the
scalar field and induced gravity on the brane, controls the initial
density and the age of the universe in this sense that these
quantities are sensitive to the proposed value of the non-minimal
coupling. For large values of $\alpha(\phi)$, the universe age and
its initial density are higher than the case with a small value of
$\alpha(\phi)$. Moreover, by increasing $\alpha(\phi)$, one of the
solutions, that is, GBIG4 disappears from the set of the solutions
(note that these results are obtained for constant values of $\psi$
and $\sigma$ while $\alpha(\phi)$ is variable). The GBIG4 branch
gives a bouncing cosmological solution. A bouncing universe goes
from an era of accelerated collapse to an expanding phase without
displaying any singularity. In the bouncing universe, the equation
of state parameter of the matter content, $\omega$,  must transit
from $\omega<-1$ to $\omega>-1$ [65]. However, current observational
data show that the equation of state parameter $\omega$ was larger
than $-1$ in the past and is less than $-1$ today [66,67]. In our
framework, we have seen that in a suitable domain of parameters
space, by increasing $\alpha(\phi)$ values the GBIG4 branch
containing a bouncing cosmology will disappear. Even for a constant
$\alpha(\phi)$ there is no bouncing solution for any values of
$\psi$. For example with $\psi=-5$, this branch disappear
completely. These arguments show that inclusion of non-minimal
coupling of scalar field and induced gravity on the brane can be
used to fine-tune braneworld cosmological models in the favor of
observational data.
\section{Time Evolution of the Branches }
In this section we discuss more general case of a time-dependent
non-minimal coupling. All arguments presented in the preceding
sections can be reconsidered in this time-dependent framework. We
focus on Minkowski bulk ($\psi=0$) with brane tension($\sigma\neq0$)
for instance. Starting with Friedmann equation (35), we try the
following ansatz
\begin{equation}
\phi(t)\propto t^{-\nu}
\end{equation}
in order to investigate the late-time behavior of this scenario. As
has been mentioned at the end of the subsection $4.2$, the values of
$\xi$ are constraint to be in the range of $\xi\leq 0.99 \phi^{2}$
and $\xi\geq 1.01 \phi^{2}$. By adapting these conditions and
choosing $\alpha(\phi)=1-\xi \phi^{2}$, equation (35) can be
rewritten as follows
\begin{equation}
(1+\gamma h^{2})^{2}h^2=\Bigg[ (1-\xi t^{-2\nu})h^{2}-\Big(\mu+
\mu'+\sigma -2\kappa_{5}^{2} \nu \xi t^{(-2\nu-1)}h\Big) \Bigg]^{2}.
\end{equation}
where proportionality constant in (46) has been set equal to unity.
The constraints on $\xi$ are now time-dependent as $\xi\geq 0.99
t^{-2\nu}$ and $\xi \leq 1.01 t^{-2\nu}$. The results of numerical
solution of this equation are shown in figures $11$ and $12$ for
different values of $\xi$. Note that three graphs of figure $11$ (
and also  figure $12$ with a different value of non-minimal coupling
coefficient $\xi$) are corresponding to three branches of figure
$6$, but now with time variation of non-minimal coupling. These
solutions are various possibilities of GBIG scenario with a time
varying non-minimally coupled scalar field on the brane. For
instance, figures $11a$ and $12a$ are corresponding to GBIG2 branch
of the scenario. On the other hand, figures $11b$ and $12b$ are
corresponding to GBIG1 branch and finally, $11c$ and $12c$ are
corresponding to GBIG3 branch. The main point to stress here is the
fact that in the presence of explicit time evolution of scalar
field, these branches show more or less the same late-time behavior
as discussed in previous sections.
\newpage
\begin{figure}[htp]
\begin{center}
\includegraphics{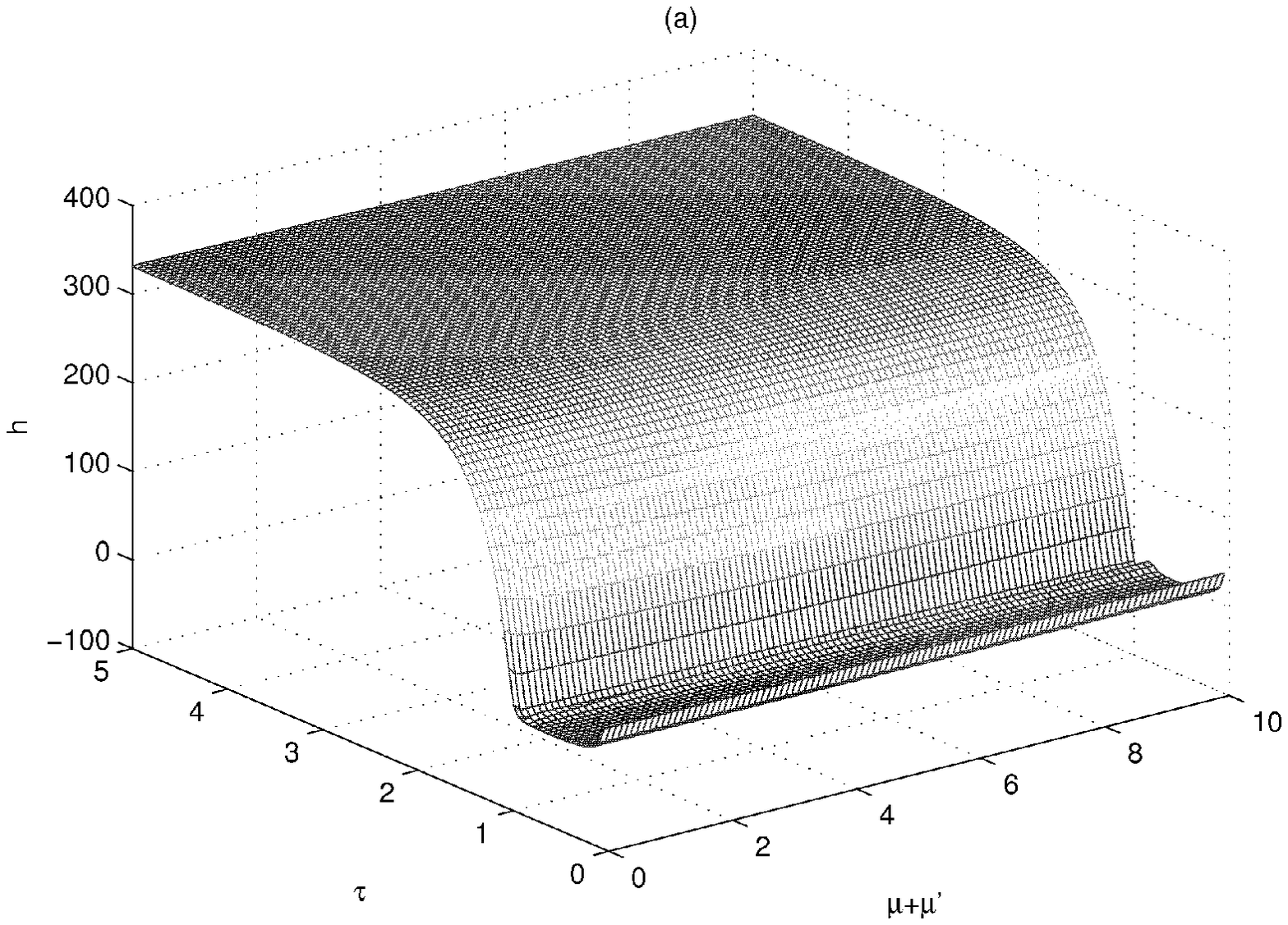} \vspace{5cm}\includegraphics{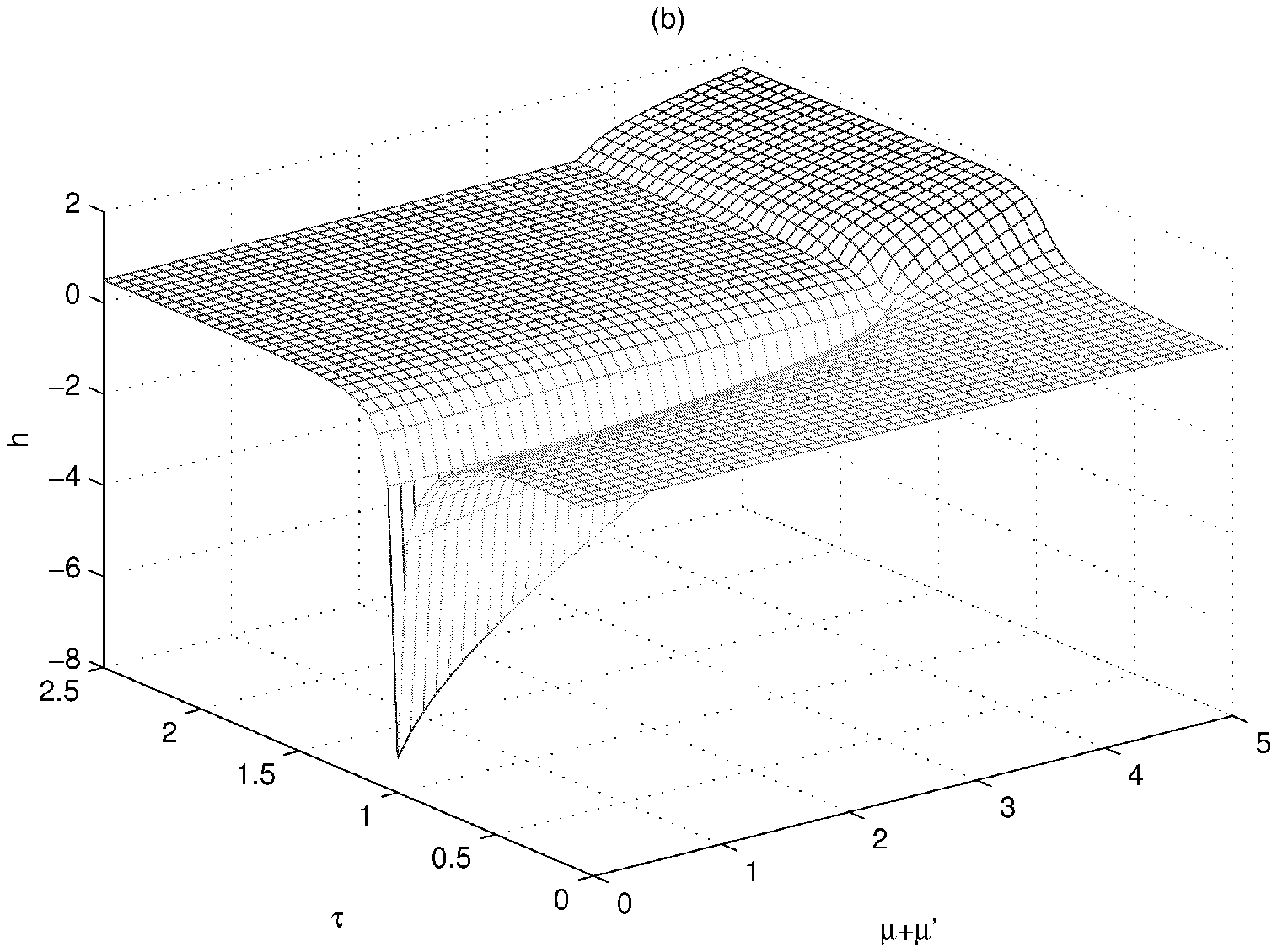}\vspace{5cm}\includegraphics{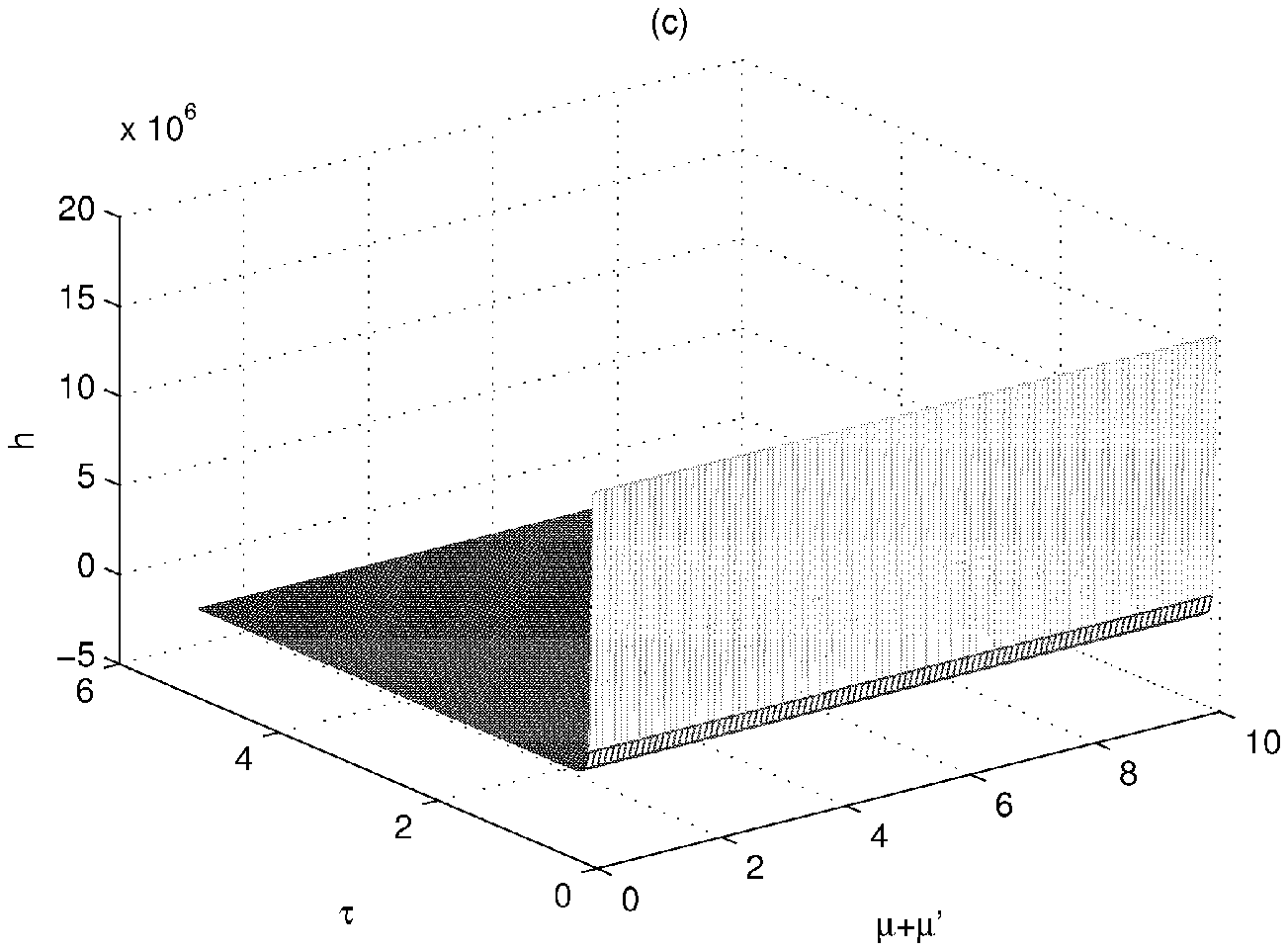}
\end{center}
\vspace{2.5cm}
 \caption{\small { Solutions of the Friedmann equation with a negative tension brane in a Minkowski
 bulk. We have assumed
 $\gamma=0.003$, $\sigma=-4$, $\psi=0$, $\nu=0.9$, $\kappa_{5}^{2}=1$, $\xi=0.99 \phi^{2}$.}}
\end{figure}

\begin{figure}[htp]
\begin{center}
\includegraphics{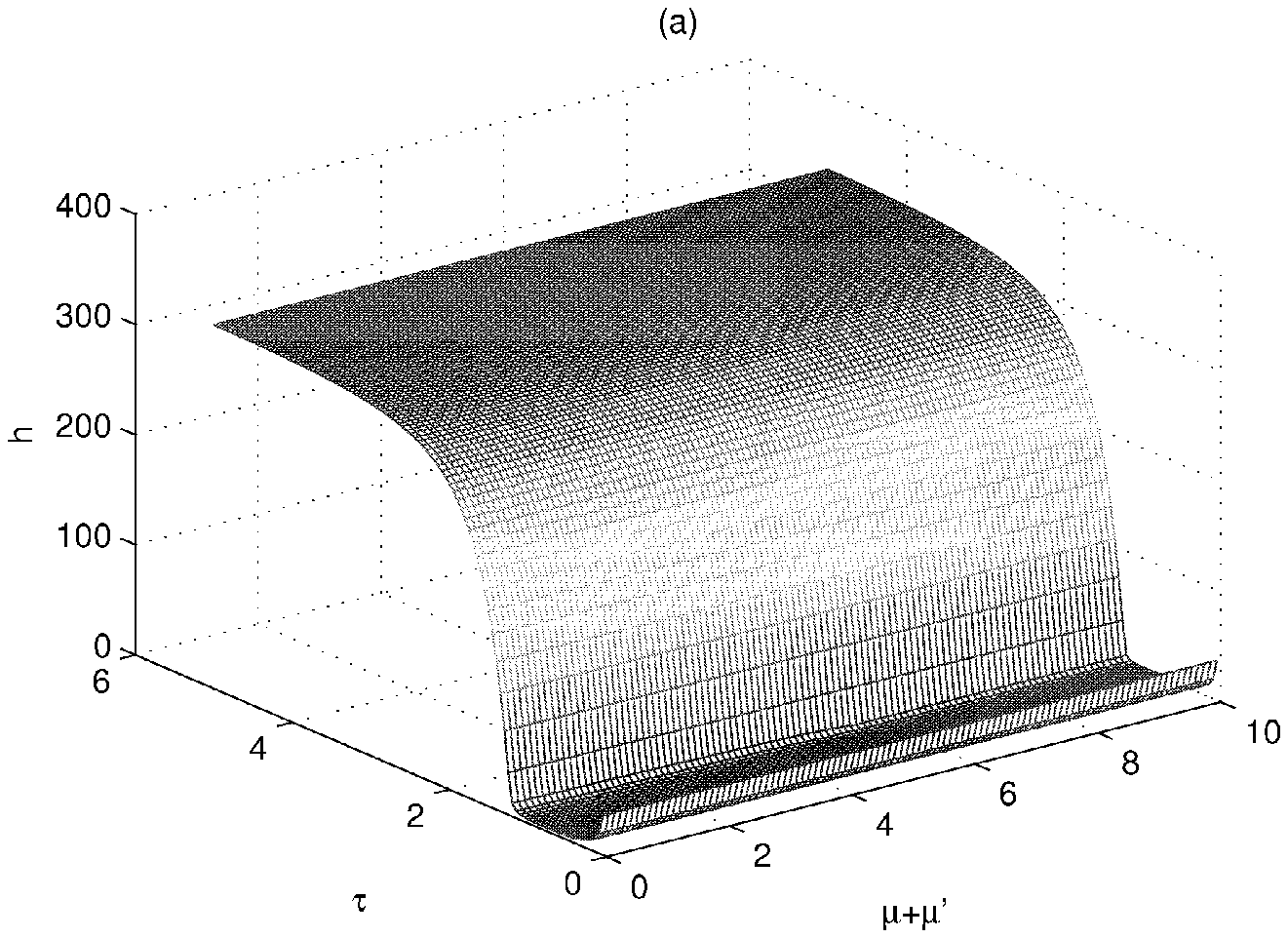} \vspace{5cm}\includegraphics{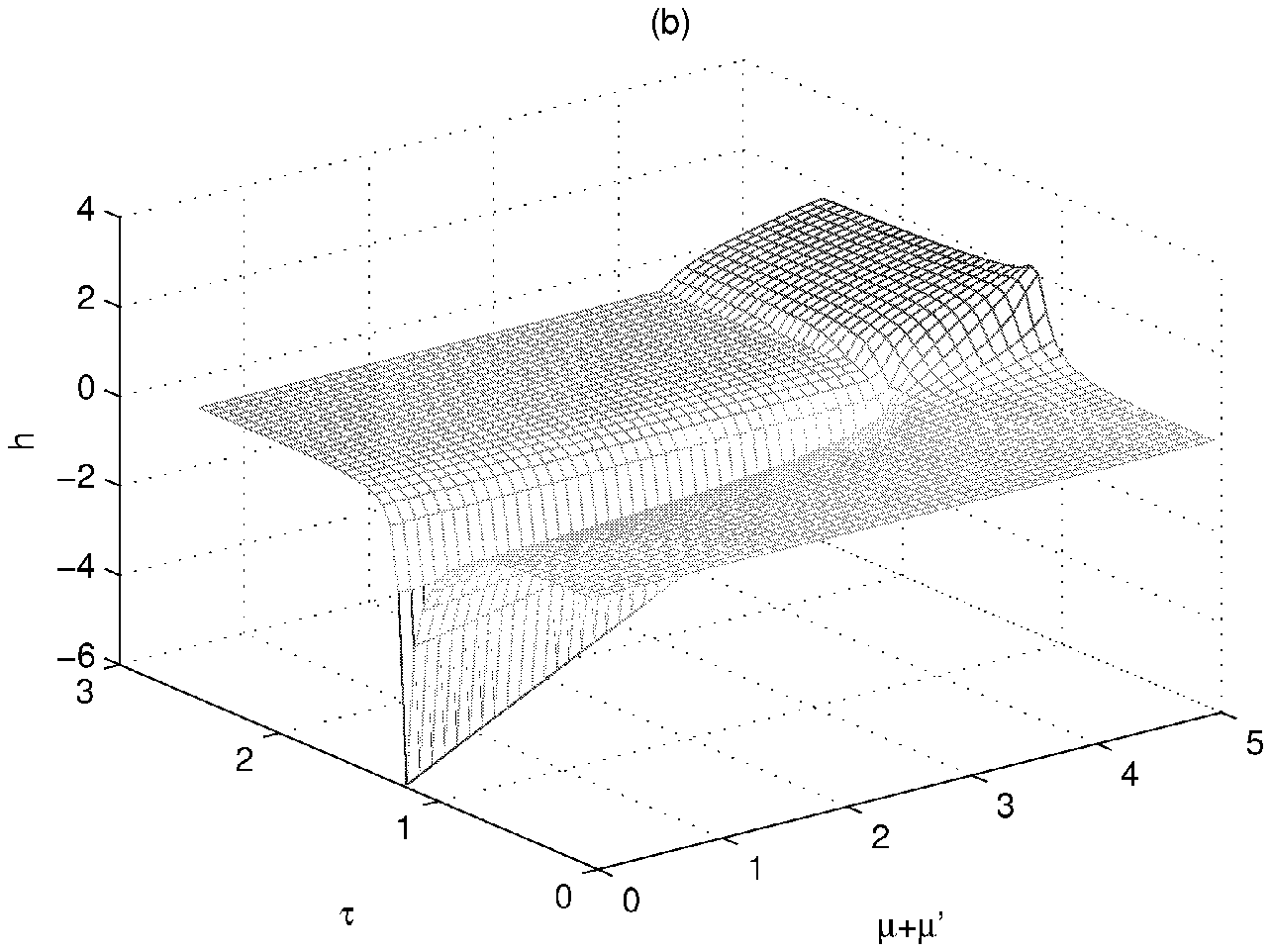}\vspace{5cm}\includegraphics{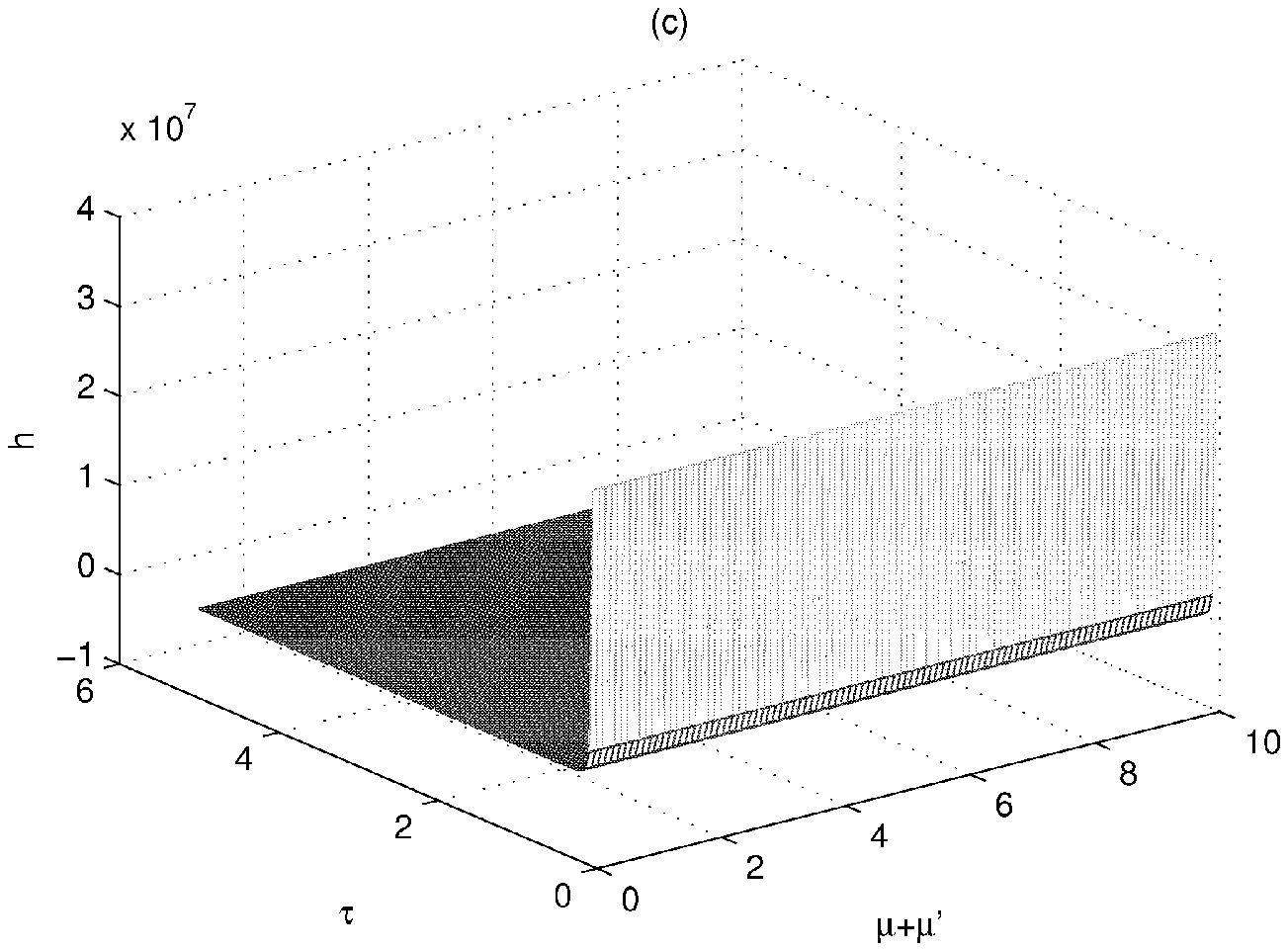}
\end{center}
 \caption{\small { Solutions of the Friedmann equation with a negative tension brane in a Minkowski
 bulk. We have assumed
 $\gamma=0.003$, $\sigma=-4$, $\psi=0$, $\nu=0.9$, $\kappa_{5}^{2}=1$, $\xi=2 \phi^{2}$.}}
\end{figure}

\newpage
Finally the issue of stability of the self-accelerated solutions
should be stressed here. It has been shown that the
self-accelerating branch of the DGP model contains a ghost at the
linearized level [68]. The ghost carries negative energy density and
it leads to the instability of the spacetime. The presence of the
ghost can be related to the infinite volume of the extra-dimension
in DGP setup. When there are ghosts instabilities in
self-accelerating branch, it is natural to ask what are the results
of solutions decay. One possible answer to this question is as
follows: since the normal branch solutions are ghost-free, one can
think that the self-accelerating solutions may decay into the normal
branch solutions. In fact for a given brane tension, the Hubble
parameter in the self-accelerating universe is larger than that of
the normal branch solutions. Then it is possible to have nucleation
of bubbles of the normal branch in the environment of the
self-accelerating branch solution. This is similar to the false
vacuum decay in de Sitter space. However, there are arguments
against this kind of reasoning which suggest that the
self-accelerating branch does not decay into the normal branch by
forming normal branch bubbles [68]. It was also shown that the
introduction of Gauss-Bonnet term in the bulk does not help to
overcome this problem [69]. In fact, it is still unclear what is the
end state of the ghost instability in self-accelerated branch of DGP
inspired setups (for more details see [68]). On the other hand,
non-minimal coupling of scalar field and induced gravity provides a
new degree of freedom which requires special fine tuning and this my
provide a suitable basis to treat ghost instability. As we have
shown, non-minimal coupling of scalar field and induced gravity has
the capability to remove bouncing solutions. It seems that this
additional degree of freedom has also the capability to provide the
background for a more reliable solution to ghost instability. This
issue deserves as a new research program.

\section{Summary and Conclusions}
DGP model modifies the IR sector of general relativity. In the UV
limit of a reliable theory, stringy effects should play important
role. In this viewpoint, to discuss both UV and IR limit of the
scenario simultaneously, the DGP model is not sufficient alone and
we should incorporate stringy effects via inclusion of the
Gauss-Bonnet terms. The presence of GB term removes the big bang
singularity, and the universe starts with an initial finite density.
Non-minimal coupling of scalar field and induced gravity on the
brane which is motivated from several compelling reasons, controls
the value of the initial density in a finite big bang cosmology on
the brane. This is not the only importance of non-minimal coupling
of scalar field and induced gravity; non-minimal coupling provides a
mechanism for generating spontaneous symmetry breaking at Planck
scale on the brane. In this respect, non-minimal coupling of scalar
field and induced gravity on the brane itself is a high energy
correction of the theory and it is natural to expect that this
effect couples with stringy effects in Planck scale. Investigation
of the late-time behavior of DGP scenario with GB and non-minimal
coupling effects provides a framework for constraining non-minimal
coupling using recent observational data. In our model, the values
of non-minimal coupling are constraint so that the values that $\xi$
can attain are constraint to be in the range of $\xi\leq
0.99\phi^{-2}$ and $\xi\geq 1.01\phi^{-2}$. In our setup, these
constraints are dependent on the scalar field dynamics and this is
reasonable since essentially dynamics of scalar field affects its
coupling to gravity. One of the main outcome of our analysis is the
implication of non-minimal coupling on bouncing cosmologies. In the
bouncing universe, the equation of state parameter of the matter
content, $\omega$, must transit from $\omega<-1$ to $\omega>-1$.
However, current observational data show that the equation of state
parameter $\omega$ was larger than $-1$ in the past and is less than
$-1$ today. In our framework, we have seen that with a suitable
parameter space, by increasing $\alpha(\phi)$ values, the GBIG4
branch containing a bouncing cosmology will disappear. Even for a
constant $\alpha(\phi)$ there is no bouncing solution for any values
of $\psi$. These arguments show that inclusion of non-minimal
coupling of scalar field and induced gravity on the brane can be
used to fine-tune braneworld cosmological models in the favor of
observational data. Although most of the arguments in the paper are
based on a time independent non-minimal coupling, but as we have
shown, inclusion of an explicit time dependence of non-minimal
coupling will not change the physical nature of the solutions.
Finally we have discussed the issue of ghost instabilities in
self-accelerated solutions and possible impacts of Gauss-Bonnet term
and non-minimal coupling on this issue.

{\bf Acknowledgment}\\
This work has been supported partially by Research Institute for
Astronomy and Astrophysics of Maragha, Iran.\\

{\bf APPENDIX A: Three Roots of Equation (44)}
$$h_{1}=\frac{A}{3(3\gamma-\alpha)}-\Big(6(3\gamma-\alpha)(2-3\gamma\psi+2\alpha\psi)-4A^{2}\Big)/$$
 $$\Bigg[3\times 2^{\frac{2}{3}}(3\gamma-\alpha) \Bigg(-216\gamma
A-1620 \gamma^{2}\psi A+ 72\alpha A+ 972\gamma\psi\alpha
A-144\psi\alpha^{2} A+16 A^{3}+$$
$$\Big[4\Big(6(3\gamma-\alpha)(2-3\gamma\psi+2\alpha\psi)-4A^{2}\Big)^{3}+(-216\gamma
A-1620 \gamma^{2}\psi A+72\alpha A+972\gamma\psi\alpha
A-144\psi\alpha^{2} A$$
$$+16
A^{3})^{2}\Big]^{\frac{1}{2}}\Bigg)^{\frac{1}{3}}\Bigg]+\frac{1}{6\times
2^{\frac{1}{3}}(3\gamma-\alpha)}\Bigg(-216\gamma A-1620
\gamma^{2}\psi A+ 72\alpha A+ 972\gamma\psi\alpha
A-144\psi\alpha^{2} A$$
$$+16 A^{3}+\Big[4\Big(6(3\gamma-\alpha)(2-3\gamma\psi+2\alpha\psi)-4A^{2}\Big)^{3}+(-216\gamma
A-1620 \gamma^{2}\psi A+72\alpha A+972\gamma\psi\alpha A$$
$$-144\psi\alpha^{2} A+16A^{3})^{2}\Big]^{\frac{1}{2}}\Bigg)^{\frac{1}{3}}$$

$$h_{2}=\frac{A}{3(3\gamma-\alpha)}-\Bigg(\Big(1+i\sqrt{3}\Big)\Big(6(3\gamma-\alpha)
(2-3\gamma\psi+2\alpha\psi)-4A^{2}\Big)\Bigg)/$$
 $$\Bigg[6\times 2^{\frac{2}{3}}(3\gamma-\alpha) \Bigg(-216\gamma
A-1620 \gamma^{2}\psi A+ 72\alpha A+ 972\gamma\psi\alpha
A-144\psi\alpha^{2} A+16 A^{3}+$$
$$\Big[4\Big(6(3\gamma-\alpha)(2-3\gamma\psi+2\alpha\psi)-4A^{2}\Big)^{3}+(-216\gamma
A-1620 \gamma^{2}\psi A+72\alpha A+972\gamma\psi\alpha
A-144\psi\alpha^{2} A$$
$$+16 A^{3})^{2}\Big]^{\frac{1}{2}}\Bigg)^{\frac{1}{3}}\Bigg]-\frac{\Big(1-i\sqrt{3}\Big)}{12\times
2^{\frac{1}{3}}(3\gamma-\alpha)}\Bigg(-216\gamma A-1620
\gamma^{2}\psi A+ 72\alpha A+ 972\gamma\psi\alpha
A-144\psi\alpha^{2} A$$
$$+16 A^{3}+\Big[4\Big(6(3\gamma-\alpha)(2-3\gamma\psi+2\alpha\psi)-4A^{2}\Big)^{3}+(-216\gamma
A-1620 \gamma^{2}\psi A+72\alpha A+972\gamma\psi\alpha A$$
$$-144\psi\alpha^{2} A+16 A^{3})^{2}\Big]^{\frac{1}{2}}\Bigg)^{\frac{1}{3}}$$

$$h_{3}=\frac{A}{3(3\gamma-\alpha)}-\Bigg(\Big(1-i\sqrt{3}\Big)\Big(6(3\gamma-\alpha)
(2-3\gamma\psi+2\alpha\psi)-4A^{2}\Big)\Bigg)/$$
 $$\Bigg[6\times 2^{\frac{2}{3}}(3\gamma-\alpha) \Bigg(-216\gamma
A-1620 \gamma^{2}\psi A+ 72\alpha A+ 972\gamma\psi\alpha
A-144\psi\alpha^{2} A+16 A^{3}+$$
$$\Big[4\Big(6(3\gamma-\alpha)(2-3\gamma\psi+2\alpha\psi)-4A^{2}\Big)^{3}+(-216\gamma
A-1620 \gamma^{2}\psi A+72\alpha A+972\gamma\psi\alpha
A-144\psi\alpha^{2} A$$
$$+16 A^{3})^{2}\Big]^{\frac{1}{2}}\Bigg)^{\frac{1}{3}}\Bigg]-\frac{\Big(1+i\sqrt{3}\Big)}{12\times
2^{\frac{1}{3}}(3\gamma-\alpha)}\Bigg(-216\gamma A-1620
\gamma^{2}\psi A+ 72\alpha A+ 972\gamma\psi\alpha
A-144\psi\alpha^{2} A$$
$$+16 A^{3}+\Big[4\Big(6(3\gamma-\alpha)(2-3\gamma\psi+2\alpha\psi)-4A^{2}\Big)^{3}+(-216\gamma
A-1620 \gamma^{2}\psi A+72\alpha A+972\gamma\psi\alpha A$$
$$-144\psi\alpha^{2} A+16
A^{3})^{2}\Big]^{\frac{1}{2}}\Bigg)^{\frac{1}{3}}$$\\
 where $A=\frac{d \alpha}{d \tau}\kappa_{4}^{2}$.

\end{document}